\documentclass[aps]{revtex4}
\usepackage{graphicx}
\usepackage{bm}
\begin{document}

\title[Twenty-five years of multifractals in fully developed turbulence]
{Twenty-five years of multifractals in fully developed turbulence: 
a tribute to Giovanni Paladin}

\author{G. Boffetta$^1$, A. Mazzino$^2$ and  A. Vulpiani$^3$}
\address{ 
$^1$Dipartimento di Fisica Generale, Universit\`a 
di Torino, CNISM and INFN via P. Giuria 1, 10125 Torino, Italy}
\address{$^2$ Dipartimento di Fisica, Universit\`a 
di Genova, CNISM and INFN, via Dodecaneso 33, 16146, Italy}
\address{$^3$ 
Dipartimento di Fisica, Universit\`a
di Roma ``La Sapienza'', CNISM and  INFN, p.le A. Moro 2, 00185 Roma, Italy }

\begin{abstract}
The paper {\it On the multifractal nature of fully developed turbulence
and chaotic systems}, by R. Benzi {\it et al.}  published in this journal 
in 1984  ( vol {\bf 17}, page 3521) has been a starting point of many 
investigations on the different faces of selfsimilarity and 
intermittency in turbulent phenomena. \\
Since then, the multifractal model has become a useful tool for
the study of small scale turbulence,
in particular for detailed predictions of  different
Eulerian and  Lagrangian  statistical properties. \\
In the occasion of the 50-th birthday of our unforgettable friend 
and colleague Giovanni Paladin (1958-1996), we review here  the  basic concepts 
and some applications of the multifractal model for turbulence.
\end{abstract}

\maketitle

\section{Introduction}

The idea of the multifractal approach to fully developed turbulence has been
introduced by Giorgio Parisi and Uriel Frisch 
during the Summer School {\it Turbulence and predictability of 
geophysical fluid dynamics} held in Varenna in June 1983 \cite{PF85}.
One of us (A.V.) had the chance to participate to that school
and then to coauthor, with R. Benzi, G. Parisi and G.Paladin, 
the paper published in this journal where the word multifractal appeared 
for the first time \cite{BPPV84}.

From a technical point of view the idea of the multifractal is basically
contained in the large deviation theory \cite{Ellis99,Varadhan03}
%\cite{HJKPS86}, 
which is an important chapter
of the probability theory. However the introduction of the multifractal 
description in the 1980's had an important role in statistical 
physics, chaos and disordered systems.
In particular to clarify in a rather net way that the
usual idea, coming from the critical phenomena, that just
few scaling exponents are relevant, is wrong, while an infinite
set of exponents is necessary for a complete characterization of the scaling
features.\\
As pioneering works which anticipated some aspects of 
the the multifractal approach to turbulence we can cite 
the lognormal theory of Kolmogorov \cite{K62}, the contributions of
Novikov and Stewart \cite{NS64} and Mandelbrot \cite{M74}.

This paper has no pretense to be a survey of the many applications
of the multifractal description in chaos, disordered systems
and natural phenomena; for general reviews on these aspects
see \cite{BS95,BP97,Meakin98}.
For a more mathematically oriented treatment see \cite{Harte01}.
Our aim is a  discussion on the the use of the multifractal methods
in the study of the scaling features of fully developed turbulence.

The paper is organized as follows.
Section~\ref{sec:2} is devoted to the introduction of the multifractal model
of turbulence and its connections with the $f(\alpha)$ vs $\alpha$
formalism introduced by Halsey {\it et al.} \cite{HJKPS86},  
and the large deviations theory \cite{Ellis99,Varadhan03}.
In Section~\ref{sec:3} we discuss the implication of multifractality
on Eulerian features, namely the statistical properties of the
velocity gradients and the existence of an intermediate dissipative
range. Section~\ref{sec:4} is devoted to the implications of the
multifractal nature of turbulence on Lagrangian statistics.
In Section 5 we present the Lagrangian acceleration statistics.
Section 6 treats the relative dispersion, in particular we discuss
the multifractal generalization of the classical Richardson theory.
Section 7 is devoted to the multifractal analysis of the dispersion
in two-dimensional convection.

%%%%%%%%%%%%%%%%%%%%%%%%%%%%%%%%%%%%%%%%%%%%%%%%%%%%%%%%%%%%%%%%%%%
%%%%%%%%%%%%%%%%%%%%%%%%%%%%%%%%%%%%%%%%%%%%%%%%%%%%%%%%%%%%%%%%%%%
\section{From Kolmogorov to multifractals}
\label{sec:2}

Let us consider the Navier-Stokes equations for an incompressible fluid:
\begin{equation}
\label{I.1}
 {\partial  {\bf v}  \over \partial t}  + ({\bf v} \cdot \nabla) {\bf v} =
  -{ 1 \over \rho} \nabla p   +\nu \Delta {\bf v}  +{\bf F} \, ,
  \qquad  \nabla \cdot {\bf v}   =0 \,\, .
\end{equation}
Because of the nonlinear structure of the equation, an analytical
treatment is a formidable task.
For instance in the  $3D$ case a theorem for the existence of global solution
for arbitrary $\nu$ is still missing. 

For a perfect fluid (i.e. $\nu=0$) and  in absence of external forces
(${\bf F}=0$), the evolution of the  velocity field is given by the 
Euler equation, which conserves the kinetic energy for smooth solutions.
In such a case, introducing an ultraviolet cutoff $K_{max}$ on the wave 
numbers, it is possible to build up an equilibrium 
 statistical mechanics simply following the standard approach used for the
 Hamiltonian statistical mechanics.
On the other hand, because of the so called dissipative 
anomaly \cite{F95,BJPV98},
in  $3D$ the limit $\nu \to 0$ is singular and cannot be interchanged with 
$K_{max} \to \infty$,  therefore the statistical mechanics  of an 
inviscid fluid has a rather limited relevance for the Navier-Stokes 
equations at very high Reynolds numbers
($Re=VL/\nu$, where $V$ and $L$ are the typical speed
and length of the system, respectively).

In addition,  mainly as a consequence of the non-Gaussian statistics,
even a systematic statistical approach, e.g. in term of closure
approximations,  is very difficult \cite{F95,BJPV98}.

In the fully developed turbulence (FDT) limit, i.e. $\nu \to 0$, and
in the presence of forcing at large scale,  one has
a non equilibrium statistical steady state, with an inertial
range of scales, where neither energy pumping nor dissipation acts, 
which shows strong departures from the equipartition \cite{F95,BJPV98}.
A simple and elegant explanation of the main statistical features 
of FDT is due to Kolmogorov \cite{MY75}: in a nutshell, it is assumed the
existence of a range of scales where the energy -- injected at the
scale $L$ -- flows down (with a cascade process, as remarked by
Richardson~\cite{R22}) to the dissipative scale $\ell_D \sim L Re^{-3/4}$,
where it is
dissipated by molecular viscosity. Since, practically, neither injection
nor dissipation takes place in the
inertial range, the only relevant quantity is the
average energy transfer rate $\bar{\varepsilon}$.
Dimensional counting
imposes  the power law dependence for the second order structure
function (SF)
\begin{equation}
\label{I.2}
S_2(\ell) = \langle \delta v_{\ell}^2 \rangle =
\langle \left(v(x+\ell)-v(x)\right)^2\rangle \propto
\bar{\varepsilon}^{2/3} \ell^{2/3}\, ,
\end{equation}
where, for the sake of simplicity, we ignore the vectorial nature of 
the velocity field.
The scaling law (\ref{I.2}) is equivalent to  a power spectrum $E(k)\propto
\bar{\varepsilon}^{2/3}k^{-5/3}$ in good agreement with the experimental
observations.
The original Kolmogorov theory (often indicated as K41) assumes 
self-similarity of the
turbulent flow. As a consequence, the scaling behavior of higher
order structure functions 
\begin{equation}
S_p(\ell)=\langle |v(x+\ell)-v(x)|^p\rangle \sim \ell^{\zeta_p}
\label{eq:1.1.3}
\end{equation}
is described by a single scaling exponent: $\zeta_p=p/3$.

\subsection{The multifractal model}
\label{sec:parisi}

The Navier-Stokes equations
are formally invariant under the scaling transformation:
$$
{\bf x} \to \lambda \, {\bf x} \,\,\,\, ,
{\bf v} \to  \lambda^h \, {\bf v} \,\,\,\, , 
 t \to \lambda^{1-h} \, t \,\,\, , \,\,\,
\nu \to \lambda^{h+1}\nu \,\, ,
$$
with $\lambda>0$ (indeed the Reynolds number $VL/\nu$ is invariant under the
above transformations).

The exponent  $h$ cannot be determined with only symmetry 
considerations, nevertheless there is a rather natural candidate:
$h=1/3$.
Such a  value of the exponent is suggested 
by the dimensional argument of the K41 and also, and more rigorously,
by the so-called ``4/5 law'', an exact relation derived by Kolmogorov
from the Navier--Stokes equations \cite{K41,F95}, which, under the
assumption of stationarity, homogeneity and isotropy, states
\begin{equation}
\label{I.3}
\langle \delta v^3_{||}(\ell)\rangle = - {4 \over 5} \bar{\varepsilon} \ell\,,
\end{equation}
where $\delta v_{||}(\ell)$ is the longitudinal velocity difference between 
two points at distance $\ell$.

We can say that the K41 theory corresponds to a global invariance
with $h=1/3$ and therefore $\zeta_p=p/3$.
This result is in  disagreement with
several experimental investigations \cite{AGHA84,F95}  which 
have shown deviations of the scaling exponents from $p/3$.
This phenomenon, which goes under the name of intermittency \cite{F95},
is a consequence of the breakdown of self--similarity and
implies that the scaling exponents cannot be determined on a simple
dimensional basis.

A simple way to modify the K41 consists in assuming that the energy dissipation
is  uniformly distributed on homogeneous fractal with dimension $D_F <3$.
This implies $ \delta v_{\ell}({\bf x}) \sim \ell^h$ with $h=(D_F-2)/3$
for ${\bf x}$ on the fractal and  $ \delta v_{\ell}({\bf x})$ non singular
otherwise.
This assumption (called absolute curdling or $\beta$- model)
gives:
\begin{equation}
\zeta_p={{D_F-2} \over 3}p+(3-D_F) \, .
\label{eq:1.1.5}
\end{equation}
Such a  prediction, with $D_F \simeq 2.83$ , is in fair agreement with the
experimental data for small values of $p$, but higher order scaling exponents
give a clear indication of a non linear behavior in $p$.

The multifractal model of turbulence \cite{PF85,PV87,F95} assumes that
the velocity has a local scale-invariance, i.e. there is not  a
unique scaling exponent $h$ such that $\delta v_{\ell} \sim \ell^h$,
but a continuous spectrum of exponents, each of which belonging to a
given fractal set. In other words, in the inertial range one has
\begin{equation}
\delta v_{\ell}({\bf x}) \sim \ell^h\,,
\label{I.4}
\end{equation}
if ${\bf x} \in S_h$, and $S_h$ is a fractal set with dimension $D(h)$ and
$h \in$ ($h_{min}$, $h_{max}$).  The probability to observe a given
scaling exponent $h$ at the scale $\ell$ is $P_{\ell}(h) \sim
\ell^{3-D(h)}$ and therefore one has
\begin{equation}
S_p(\ell) = \langle |\delta v_{\ell}|^p\rangle \sim
\int_{h_{min}}^{h_{max}} \ell^{h p} \ell^{3-D(h)} {\rm d} h 
\sim \ell^{\zeta_p}\,.
\label{I.5}
\end{equation}
%{\bf Sarebbe forse meglio scrivere il tutto in $\ell/L$}
For $\ell \ll 1$, a steepest descent estimation gives
\begin{equation}
\zeta_p = \min_{h}\left\{hp+3-D(h)\right\} = h^*p + 3 - D(h^*)
\label{I.6}
\end{equation}
where $h^*=h^*(p)$ is the solution of the equation $D'(h^*(p))=p$.
The Kolmogorov ``4/5'' law (\ref{I.3})
imposes $\zeta_3=1$ which implies that
\begin{equation}
D(h) \le 3 h + 2  \,,
\label{I.8}
\end{equation}
with the equality realized by $h^*(3)$.  The Kolmogorov similarity
theory corresponds to the case of only one singularity exponent
$h=1/3$ with $D(h=1/3)=3$.

Of course the computation of $D(h)$, or equivalently $\zeta_p$,
from the NSE is not at present an attainable goal.
A first step is a phenomenological approach using multiplicative
processes. Let us briefly remind the so called random $\beta$- model \cite{BPPV84}.
This model describes the energy cascade in real space looking at eddies
of size $\ell_n=2^{-n} L$, with $L$ the length at which the energy 
is injected. At the $n$- th step of the cascade a mother eddy of
size $\ell_n$ splits into daughter eddies of size $\ell_{n+1}$, and 
the daughter eddies cover a fraction $\beta_j$ ($0<\beta_j<1$) of the 
mother volume. 
As a consequence of the fact that the energy transfer is constant throughout
the cascade one has for 
 the velocity differences
$v_n=\delta v_{\ell_n}$ on scale $\ell_n$ is
non negligible only on a fraction of volume $\prod_j \beta_j$,
and it is given by
\begin{equation}
v_n=v_0 \ell_n^{1/3}\prod_{j=1}^n \beta_j^{-1/3}
\label{eq:vn}
\end{equation}
where the $\beta_j$'s are independent, identically distributed random 
variables.
Phenomenological arguments suggest: $\beta_j=1$ with probability
$x$ and $\beta_j=B=2^{-(1-3h_{min})}$ with probability $1-x$
The above multiplicative process generates a two-scale Cantor set,
which is a rather common structure in chaotic systems.
The scaling exponents are:
\begin{equation}
\zeta_p={p \over 3} -\ln_2[x+(1-x)B^{1-p/3}] \,\,
\label{eq:rbm}
\end{equation}
corresponding to
\begin{equation}
D(h)=3+ \bigl(3h-1 \bigr) 
\Bigl[ 1+\ln_2\Bigl({ {1-3h} \over {1-x}} \Bigr)\Bigr] 
+ 3 h \ln_2 \left({x \over 3h} \right) \, .
\end{equation}
The two limit cases are $x=1$, i.e. the K41, and $x=0$
which is the $\beta$- model with $D_F=2+3 h_{min}$.
Using $x=7/8$, $h_{min}=0$ (i.e. $B=1/2$) one has a good fit for the
$\zeta_p$ of the experimental data at high Reynolds numbers.

%%%%%%%%%%%%%%%%%%%%%%%%
\begin{figure}[!ht]
\centering
\includegraphics[width=10 cm]{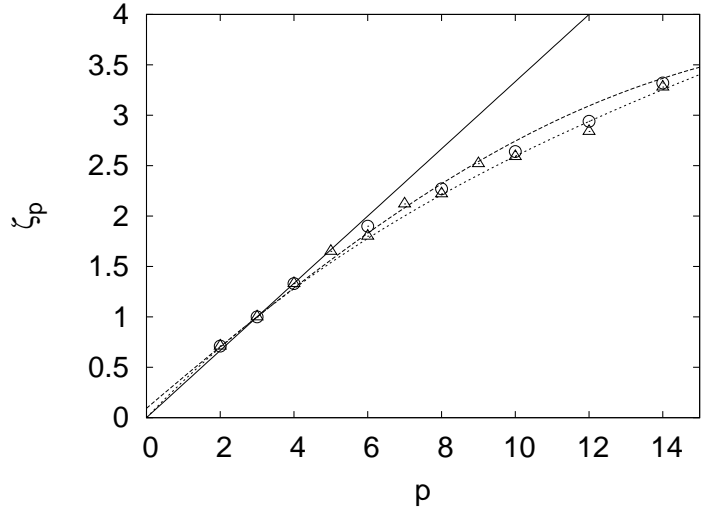}
\caption{Structure function scaling exponents $\zeta_p$
plotted vs $p$. Circles and triangles correspond to the data of
Anselmet at al. \cite{AGHA84}. The solid line corresponds to Kolmogorov
scaling $p/3$; the dashed line is the random beta model prediction
(\ref{eq:rbm}) with $B=1/2$ and $x=7/8$; the dotted line is 
the She-Leveque prediction (\ref{eq:sl}) with $\beta=2/3$.}
\label{fig:1}
\end{figure}
%%%%%%%%%%%%%%%%%%%%%%%%

Of course it is not so astonishing to find a model to fit
the experimental data. Indeed, there are now many phenomenological
models for $D(h)$ which provide scaling exponents in 
agreement with experimental data. A popular one is the so-called
She-Leveque model \cite{SheLeveque94} which is reproduced by the multifractal
model with 
$$
D(h)=1+{2 \beta -3 h-1 \over \ln \beta}
\left[ 1 - \ln \left({2 \beta - 1 - 3 h \over 2 \ln \beta}\right)\right]
$$
and gives for the scaling exponents
\begin{equation}
\zeta_p = {2 \beta-1 \over 3} p + 2(1-\beta^{p/3})
\label{eq:sl}
\end{equation}
which are close to the experimental data for $\beta=2/3$.
Another important model, which was introduced by Kolmogorov 
himself without reference to the multifractal model, is the 
log-normal model which will be discussed in Section~\ref{sec:1.3}.

The relevance, and the success, of the multifractal approach
is in the possibility to predict, and test, nontrivial
statistical features, e.g. the pdf of the velocity gradient, the
existence of an intermediate dissipative range and precise
scaling for Lagrangian quantities.
Once $D(h)$ is obtained by a fit of the experimental data,
e.g. from the $\zeta_p$, then all the predictions obtained
in the multifractal model framework must be verified without
additional free parameters. 

%%%%%%%%%%%%%%%%%%%%%%%%%%%%%%%%%%%%%%%%%%%%%%%%%%%%%%%%%%%%%%%%%%%
%%%%%%%%%%%%%%%%%%%%%%%%%%%%%%%%%%%%%%%%%%%%%%%%%%%%%%%%%%%%%%%%%%%
\subsection{Relation between the original multifractal model
and the $f(\alpha)$ vs $\alpha$}
\label{sec:1.3}

The multifractal  model for the FDT previously discussed is 
linked to the so called  $f(\alpha)$ vs $\alpha$ description
of the singular measures (e.g. in chaotic attractors) 
\cite{HJKPS86,BJPV98,AFLV92}.
In order to show this connection let us recall the Kolmogorov revised 
theory \cite{K62} (called K62) stating that the velocity increments
$\delta v_{\ell}$ scales as $(\epsilon_{\ell} \ell)^{1/3}$ where
$\epsilon_{\ell}$ is the energy dissipation space-averaged over a cube of edge
$\ell$.
Let us introduce the measure 
$\mu({\bf x})=\epsilon({\bf x})/{\bar \epsilon}$,
a partition of non overlapping cells of size $\ell$ 
and the coarse graining probability 
$$
P_i(\ell)=\int_{\Lambda_{\ell}({\bf x})} d\mu({\bf y})
$$
where $\Lambda_l({\bf x}_i)$ is a cube of edge $\ell$ centered in ${\bf x}_i$,
of course $\epsilon_{\ell} \sim \ell^{-3} P(\ell)$.
Denoting with $\alpha$ the scaling exponent of $P_{\ell}$ 
and with $f(\alpha)$ the fractal dimension of the subfractal with
scaling exponent $\alpha$, we can introduce the Renyi dimensions
$d_p$:
$$
\sum_i P_i(\ell)^p \sim \ell^{(p-1)d_p}
$$
where the sum is over the non empty boxes.
A simple computation gives
$$
(p-1)d_p=\min_{\alpha}[p \alpha -f(\alpha)] \,\, .
$$
Noting that 
$
<\epsilon_{\ell}^p>=\ell^3 \sum\epsilon_{\ell}^p 
$
we have
$$
<\epsilon_{\ell}^p>\sim \ell^{(p-1)(d_p-3)}
$$
therefore one has the correspondence
$$
h \leftrightarrow {{\alpha -2} \over 3} \,\,\, , \,\,\, 
D(h) \leftrightarrow f(\alpha)\,\,\, , \,\,\, 
\zeta_p={p \over 3}+ \bigl({p \over 3} -1\bigr)
\bigl(d_{p \over 3}-3\bigr) \,\, .
$$
Of course the result $\zeta_3=1$, once assumed 
$\delta v \sim (\epsilon_{\ell} \ell)^{1/3}$, holds for any $f(\alpha)$.
Let us note that the lognormal theory K62 where
$$
\zeta_p= {p \over 3} +{\mu \over 18}p(3-p)
$$
is a special case of the multifractal model, 
where there are not restriction on the values of $h$ and  $D(h)$ 
is a parabola with a maximum at $D_F=3$:
$$
D(h) = - {9 \over 2 \mu} h^2 + {3 \over 2}(2+\mu) h 
- {4 -20 \mu + \mu^2 \over 8 \mu}
$$
and the parameter $\mu$ is determined by the fluctuation of
$\ln \epsilon_{\ell}$.

%%%%%%%%%%%%%%%%%%%%%%%%%%%%%%%%%%%%%%%%%%%%%%%%%%%%%%%%%%%%%%%%%%%
\subsection{A technical remark on multifractality}
\label{sec:2.4}
To obtain the scaling behavior of $S_p(\ell)\sim (\ell/L) ^{\zeta_p}$
given by (\ref{I.5}) 
with $\zeta_p$ obtained from (\ref{I.6}), 
one has to assume that the exponent $ph +3 -D(h) $ has a minimum, $\zeta_p$, 
which is a function of $h$,  and that such an exponent behaves 
quadratically with $h$ in the vicinity of the minimum. This is the basic 
assumption to apply the Laplace's method of steepest descent \cite{BO99}.
The point we would like to recall here is that, for small separations, $\ell$,
it is true that $S_p(\ell)\sim (\ell/L) ^{\zeta_p}$ 
but with a logarithmic prefactor:
\begin{equation}
S_p(\ell ) \sim \left [-\ln \left (\frac{\ell}{L}\right )\right ]^{-1/2}
\left ( \frac{\ell}{L}\right )^{\zeta_p}
\end{equation}
 Such a prefactor is usually not considered in the
naive application of Laplace method leading to  (\ref{I.5}).
The presence of such logarithmic correction, if present, 
would clearly invalidate the
$4/5$-th law (\ref{I.3}), one of the very few exact results in fully developed 
turbulence.\\
The question on whether such logarithmic correction is likely has
quantitatively  been
addressed by Frisch et al. \cite{FMMY05}. There, exploiting the 
refined large-deviations theory, the Authors were able to explain in which 
way the logarithmic contribution cancels out thus giving rise to a prediction
fully compatible with the naive (a priory unjustified) procedure to extract the scaling behavior  (\ref{I.5}). The key
point is that the leading order large deviation result for
the probability $P_{\ell }(h)$ to be within
 a distance $\ell$ of the set carrying singularities of scaling
 exponent between $h$ and $h+ d h$,
\begin{equation}
  P_{\ell}(h)\sim \left(\frac{\ell}{L}\right)^{3-D(h)}\; ,
% \left[-\ln\frac{\ell}{L}\right]^{1/2} d h\;,
 \end{equation}
must be generalized to take into account next subleading order.
In doing so, as a result  one obtains  \cite{FMMY05}
\begin{equation} \label{better}
 P_{\ell}(h)\sim \left(\frac{\ell}{L}\right)^{3-D(h)}\left[-\ln\frac{\ell}{L}\right]^{1/2},
 \end{equation}
which contains subleading logarithmic correction.
It is worth observing that despite the multiplicative character 
of the logarithmic correction one speaks of ``subleading correction''.
This is justified by the fact that  the correct statement of the 
large-deviations leading-order result 
involves the logarithm of the probability divided
by the logarithm of the scale. The correction is then a subleading additive
term.\\
Once the expression (\ref{better}) is plugged in the integral
\begin{equation}
S_p(\ell ) \sim \int dh P_{\ell}(h) \left ( \frac{\ell}{L}\right )^{p h}
\end{equation}
and the saddle point estimation is carried out according to \cite{BO99},
logarithms disappear and  the expected $4/5$-th law emerges.\\
It is worth mentioning that the presence of a square root of a logarithm
 correction in the multifractal probability density had already been
 proposed by \cite{MS89} on the basis of a normalization requirement.
In that paper, the Authors
observed that without such a correction the singularity spectrum
 $f(\alpha)$ comes out wrong; they also pointed out that a similar correction
 has been proposed by \cite{vdWS88} in connection with the measurement of
 generalized Renyi dimensions.

%%%%%%%%%%%%%%%%%%%%%%%%%%%%%%%%%%%%%%%%%%%%%%%%%%%%%%%%%%%%%%%%%%%
%%%%%%%%%%%%%%%%%%%%%%%%%%%%%%%%%%%%%%%%%%%%%%%%%%%%%%%%%%%%%%%%%%%
\section{Implications of multifractality on Eulerian features}
\label{sec:3}

At first we note that a consequence of presence  of the intermittency 
the Kolmogorov scale does not take a unique value.
The local dissipative scale $\ell_{D}$
is determined by imposing the effective Reynolds number to be of order
unity:
\begin{equation}
Re(\ell_D)={\delta v_{D} \ell_{D} \over \nu} \sim 1\,,
\label{II.1}
\end{equation}
therefore the dependence of $\ell_{D}$ on $h$ is thus
 \begin{equation}
\ell_{D}(h) \sim L Re^{-{1 \over 1+h}}
\label{II.2}
\end{equation}
where $Re=Re(L)$ is the large scale Reynolds number \cite{PV_PRA87}.

In this section we will show that
the fluctuations of the dissipative scale, due to  
the intermittency in the turbulent cascade, is relevant of the statistical
features of the  velocity differences and velocity gradient,
and in addition it
implies the
existence of an intermediate region between the inertial and
dissipative range \cite{FV91}.

%%%%%%%%%%%%%%%%%%%%%%%%%%%%%%%%%%%%%%%%%%%%%%%%%%%%%%%%%%%%%%%%%%%
\subsection{The Pdf of the velocity differences and velocity gradient}
\label{sec:3.1}
\begin{figure}[!ht]
\centering
\includegraphics[width=10 cm]{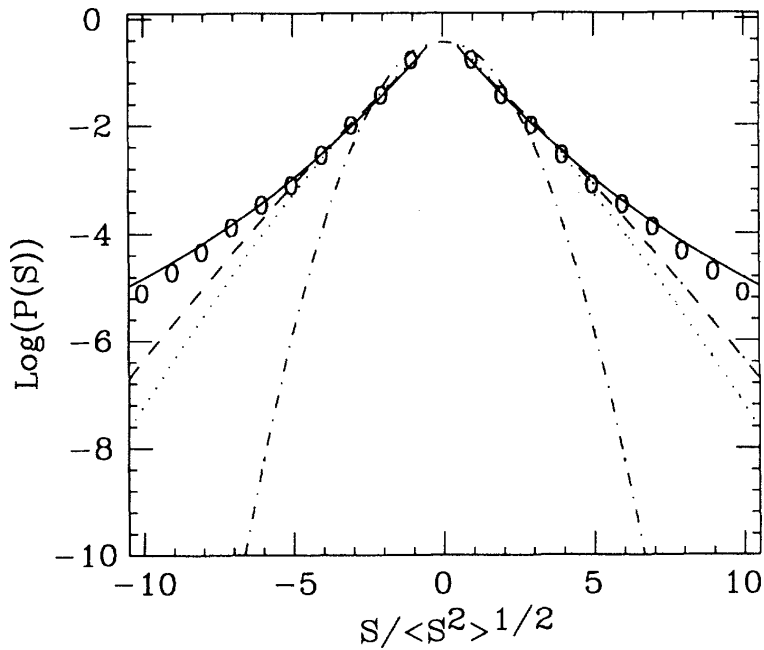}
\caption{Log-linear plot of the PDF of velocity gradients
$s$ rescaled with the rms value. Points represent experimental 
data from \cite{VM91}, solid line is the multifractal prediction
with the random beta model, dotted and dashed lines 
represent the K41 and beta model results respectively.}
\label{fig:2}
\end{figure}

Let us denote by $s$ the longitudinal velocity gradient. Such a quantity 
can immediately
be expressed in terms of the singularity exponents $h$ as
\begin{equation}
|s| \sim \frac{\delta v_{\ell_D}}{\ell_D} = v_0 \ell_D^{h-1}=
v_0^{\frac{2}{1+h}}\nu^{\frac{h-1}{h+1}}
\label{eq:s}
\end{equation}
where we used the fact that $\delta v_{\ell } \sim v_0 l^h$ from (\ref{I.4})
and we have exploited (\ref{II.2}).
From (\ref{eq:s}) we realize that we can easily express the probability density 
function (PDF) of $s$ (for a fixed $h$), $P_h(s)$, 
in terms of the PDF, $\Pi(V_0)$, 
of the large-scale velocity
differences $V_0$, with $v_0\equiv|V_0|$. 
The latter PDF is indeed known to be accurately described
by the Gaussian distribution \cite{FS91}.
The link between the two PDFs is given by the standard relation:
\begin{equation}
P_h(s)=\Pi(V_0) \left |  \frac{d V_0}{d s} \right |
\end{equation}
from which one immediately gets:
\begin{equation}
P_h(s) \sim \frac{\nu}{|s|}^{\frac{1-h}{2}}e^{-\frac{\nu^{1-h}|s|^{1+h}}{2\langle 
V_0^2\rangle }}  .
\end{equation}
The K41 theory and the $\beta$ model correspond to $h=1/3$
and $h=(D_F-2)/3$, respectively. In both cases, a stretched exponential form
for the PDF is predicted with an exponent, $1+h $,  larger than one.
Experimental data (see e.g.~\cite{CGH90,VM91}) 
are not consistent with such a prediction
being actually compatible with a stretched exponent whose value is 
smaller than one.\\
The multifractal description has thus to be exploited to capture
those experimental evidences. To do that we recall the expression 
(\ref{eq:vn}) for the random $\beta$ model 
\begin{equation}
v_n=v_0 \ell_n^{1/3}\prod_{j=1}^n \beta_j^{-1/3}
\end{equation}
from which the probability distribution of the velocity increments $v_n$
reads:
\begin{equation}
P(v_n)=\int \Pi(V_0) dV_0\int\delta(v_n-v_0 l_n^{1/3}\prod_{i=1}^n\beta_i^{-1/3})
\prod_{j=1}^n \beta_i \mu(\beta_i) d\beta_i .
\end{equation}
Here $\mu (\beta_i) $ is the probability density of the $\beta_i$'s
assumed to be of the form:
\begin{equation}
\mu (\beta_i) = x \delta(\beta_i-1) + (1- x ) \delta(\beta_i-B)
\end{equation}
with $B=2^{-(1-3h_{min})}$. \\
Since the $\beta_i$'s are identically distributed, the above integral 
becomes:
\begin{equation}
P(v_n)=\sum_{K=0}^n {n\choose K} x^{n-K} (1-x)^K B^{4 K/3}l_n^{-1/3}
e^{-C B^{2 K/3}l_n^{-2/3}v_n^2}
\label{eq:pdfdv}
\end{equation}
with $C\equiv (2\langle V_0^2\rangle)^{-1}$.
It is easy to see \cite{BBPVV91} the passage of the above PDF from a Gaussian form at large scales (small $n$)  to an exponential-like form at small scales
(large $n$).\\
To obtain the gradient PDF from (\ref{eq:pdfdv}) it is sufficient to stop
the sum at $n=N$ such that $v_N l_N/\nu=1$. This is equivalent to say
\begin{equation}
l_N^2=2^{-2N}\sim \frac{\nu}{s}\qquad\mbox{or}\qquad 
N=\frac{\ln \frac{s}{\nu}}{2\ln 2} .
\end{equation}
By noting that $B^{2N} = (\nu/s)^{1-3 h_{min}}$
the resulting gradient PDF reads:
\begin{equation}
P(s)\sim \sum_{K=0}^N {N\choose K} x^{N-K} (1- x )^K 
\left( \frac{\nu}{|s|}\right )^{(1+2 q)/3}e^{-C \nu^{(2+q)/3} |s|^{(4-q)/3}}
\end{equation}
where $q\equiv K(1-3 h_{min})/N$. The K41 prediction corresponds to considering 
only the term $K=0$ with $x =1 $.\\
We already discussed that $x = 7/8$ and $h_{min}=0$ provides
a good fit for the scaling exponents $\zeta_p$ of the structure functions
in the limit of high Reynolds numbers. The same parameters give 
a PDF behavior in good agreement with available experimental data
(see \cite{BBPVV91} and Figure \ref{fig:2}).

%%%%%%%%%%%%%%%%%%%%%%%%
%%%%%%%%%%%%%%%%%%%%%%%%

%%%%%%%%%%%%%%%%%%%%%%%%%%%%%%%%%%%%%%%%%%%%%%%%%%%%%%%%%%%%%%%%%%%%%%
\subsection{Intermediate dissipative range}
\label{eq:3.2}

Let us now show, that as consequence 
of the fluctuations of the dissipative scale one has  the
existence of an intermediate region (the Intermediate Dissipative Range, IDR) 
between the inertial and
dissipative range \cite{FV91}.
The presence of  fluctuations of $\ell_{D}$, see (\ref{II.2}),
modifies the evaluation of the structure functions
(\ref{I.5}): for a given $\ell$, the saddle point evaluation
remains unchanged if, for the selected exponent
$h^*(p)$, one has $\ell_{D}(h^*(p)) < \ell$.  If, on the contrary, the
selected exponent is such that $\ell_{D}(h^*(p)) > \ell$ the saddle
point evaluation is not consistent, because at scale $\ell$ the
power--law scaling (\ref{I.4}) is no longer valid.  In this
intermediate dissipation range the integral in
(\ref{I.5}) is dominated by the smallest acceptable scaling
exponent $h(\ell)$ given by inverting (\ref{II.2}), and the
structure function of order $p$ a pseudo--algebraic behavior, i.e. a
power law with exponent $p h(\ell) + 3 - D(h(\ell))$ which depends on
the scale $\ell$.  Taking into account the fluctuations of the
dissipative range \cite{FV91}, one has for the structure functions
\begin{equation}
S_p(\ell) \sim \left\{
\begin{array}{lll}
\ell^{\zeta_p} & \mbox{if} & \ell_{D}(h^*(p)) < \ell \\
\ell^{h(\ell) p + 3 -D(h(\ell))} & \mbox{if} & \ell_{D}(h_{min}) < \ell < 
\ell_{D}(h^*(p))\,.
\end{array}
\right.
\label{II.3}
\end{equation}

A simple calculation \cite{FV91,F95} shows that it is possible to find
a universal description valid both in the inertial and in the
intermediate dissipative ranges.  Let us discuss this point for the
energy spectrum $E(k)$.  Introducing the rescaled variables
\begin{equation}
F(\theta)={\ln E(k) \over \ln Re} \qquad {\mbox {and}}\qquad 
\theta={\ln k \over \ln Re} 
\label{II.4}
\end{equation}
one obtains the following behavior
\begin{equation}
F(\theta) = \left\{
\begin{array}{lll}
-(1+\zeta_2) \theta & \mbox{for} & \theta < {1 \over 1+h^*(2)} \\
-2 -2 \theta + \theta D(\theta^{-1}-1) & \mbox{for} &
{1 \over 1+h^*(2)} < \theta < {1 \over 1+h_{min}}
\end{array}
\right.
\label{II.5}
\end{equation}
The prediction of the multifractal model is that $\ln E(k)/\ln Re$ is
an universal function of $\ln k/\ln Re$. This is in contrast with the
usual scaling hypothesis according which $\ln E(k)$ should be a
universal function of $\ln (k/k_D)$). The multifractal universality has
been tested by collapsing energy spectra obtained from turbulent flow
in a wide range of $Re$ \cite{GC91}, see also \cite{BCVV99}.

%%%%%%%%%%%%%%%%%%%%%%%%%%%%%%%%%%%%%%%%%%%%%%%%%%%%%%%%%%%%%%%%%%%%%%%
\subsection{Exit times for turbulent signals and the IDR}
\label{sec:2.3}

In the following we will discuss a method  alternative 
to the study of the structure functions
which allows for a deeper understanding of the IDR.

Basically in typical experiments one is forced
to analyze one-dimensional string of data $v(t)$, e.g. the output of hot-wire
anemometer, and the Taylor Frozen-Turbulence Hypothesis
is used to bridge measurements in space with measurements in time.
As a function of time increment, $\tau$, structure functions
assume the form: $ S_p(\tau) = <\!\!\left[ (v(t+\tau) -v(t)\right]^p\!\!>$.  
In the inertial range, 
$\tau_D \ll \tau \ll T_0$ (where 
$T_0 = L_0/V_0$, and the dissipative time, $\tau_D = \ell_D/V_0$)
the structure functions develop an
anomalous scaling behavior: $S_p(\tau) \sim \tau^{\zeta_p}$,
where $\tau \sim \ell/V_0$.

The main idea, which can be applied both to experimental and synthetic data,
is to take a time sequence $v(t)$, and to analyze the statistical properties
 of the exit times 
from a set of defined velocity-thresholds.
More precisely, given a reference initial time $t_0$
with velocity $v(t_0)$,  we define $\tau(\delta v)$ as the
first time necessary to have  an absolute  variation equal
 to $\delta v$ in
 the velocity data, i.e. $|v(t_0)-v(t_0 +\tau(\delta v))| = \delta v$. 
By scanning the whole
time series we  recover  the probability density functions of
$\tau(\delta v)$ at varying
 $\delta v$ from the typical 
large scale values down to  the smallest dissipative values.
Positive moments of $\tau(\delta v)$ are   dominated by 
events with a smooth velocity field, i.e. laminar bursts in
the turbulent cascade.
Let us define the Inverse Structure Functions (Inverse-SF) as \cite{BCVV99,J99}:
\begin{equation}
   \Sigma_{p}(\delta v) \equiv < \!\!\tau^p(\delta v)\!\!> \,.
   \label{II.6}
\end{equation}
It is necessary to perform  weighted  average over
the time-statistics in a weighted way. This is due to the fact that by looking
at the exit-time statistics we are not sampling the time-series
uniformly, i.e. the higher the value of $\tau(\delta v)$ is, 
the longer it remains  detectable in the time series.\\
It is possible to show \cite{BCFV02}
that the sequential time average of any observable, ${\cal A}$, 
based on exit-time
statistics, $\langle {\cal A}\rangle_e$,
is connected to the uniformly-in-time multifractal average by the relation:
\begin{equation}
\label{II.7.berti}
\langle A \rangle =
\frac{\langle A \tau  \rangle_e}{\langle \tau \rangle_e}\,.
\end{equation} 
For ${\cal A}= \tau^p(\delta v)$ the above relations becomes:
\begin{equation}
\label{II.7}
\langle \tau^p(\delta v)\rangle =
\frac{\langle \tau^{p+1} \rangle_e}{\langle \tau \rangle_e}\,.
\end{equation} 
According to the multifractal description we assume that, for 
velocity thresholds corresponding to inertial range values of the velocity
differences the following dimensional relation is valid: 
$$ \delta_{\tau} v \sim \tau^h \;\;\rightarrow \;\; \tau(\delta v) \sim \delta 
v^{1/h} \,,$$ 
and the probability to observe a value $\tau$ for the exit time
is given by inverting the multifractal probability, i.e. 
$ P(\tau \sim \delta v^{1/h} ) \sim \delta v^{[3-D(h)]/h} $.
With this ansatz in the inertial range one has:
\begin{equation}
 \Sigma_p(\delta v) \sim \int_{h_{\mathrm {min}}}^{h_{\mathrm {max}}} dh \; 
\delta v^{[p + 3-D(h)]/h}  \sim \delta v^{\chi_p} 
\label{II.8}
\end{equation}
where  with the Laplace method one obtains:  
\begin{equation}
\chi_p = \min_h\left\{[p + 3-D(h)]/h\right\}\,.
\label{II.9}
\end{equation}
Let us now consider the  IDR properties. \\
For each $p$, the saddle point evaluation 
%(\ref{saddle}) 
selects a particular $h=h_{s}(p)$ where the minimum
is reached. Let us also remark that from (\ref{II.8}) we have an
estimate for the minimum value assumed by the velocity 
in the inertial range given a certain singularity $h$:
$v_m(h) = \delta_{\tau_d(h)} v \sim \nu^{h/(1+h)}$. 
Therefore, the smallest velocity value at which the scaling (\ref{II.8})
still holds depends on both $\nu$ and $h$. 
Namely, $\delta v_{m}(p) \sim \nu^{h_{s}(p)/1+h_{s}(p)}$. The most
important  consequence is that for $\delta v < \delta v_{m}(p)$ the integral
(\ref{II.8}) is not any more dominated by the saddle point value but
by the maximum $h$ value still dynamically alive at that velocity difference,
$1/h(\delta v) = -1 -\log(\nu)/\log(\delta v)$. 
This leads for $ \delta v < \delta v_{m}(p)$ to  a pseudo-algebraic law:
\begin{equation}
\Sigma_p(\delta v) \sim
 \delta v^{{\textstyle [p+3-D(h(\delta v))]/h(\delta v)}}\,.
\label{II.10}
\end{equation}
The presence of this $p$-dependent velocity range, 
intermediate between the inertial range,
$\Sigma_p(\delta v) \sim \delta v^{\chi_p}$, and the dissipative
scaling, $\Sigma_p(\delta v) \sim \delta v^{p}$, is the IDR signature. 
\vspace{1.0truecm}

%%%%%%%%%%%%%%%%%%%%%%%%
\begin{figure}[!ht]
\centering
\includegraphics[width=10 cm]{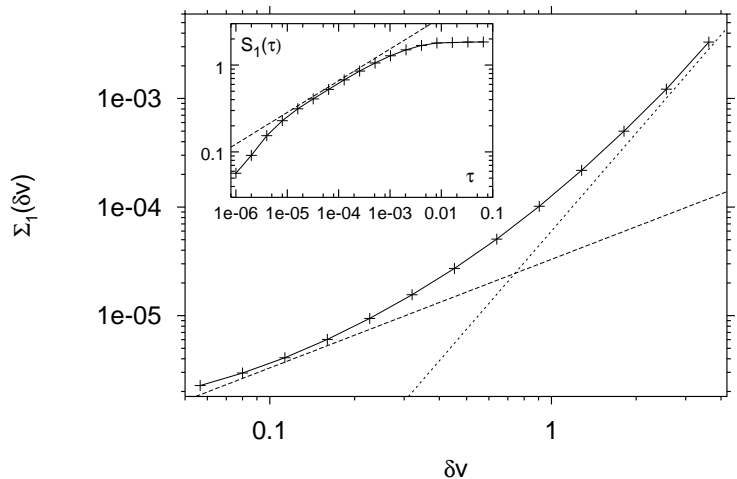}
\caption{Inverse Structure Functions $\Sigma_1(\delta v)$.
The straight lines shows the dissipative 
range behavior (dashed) $\Sigma_1(\delta v) \sim  \delta v$, 
and the inertial range non intermittent behavior (dotted)
$\Sigma_1(\delta v)\sim  (\delta v)^3$. 
The inset shows the direct structure function $S_1(\tau)$ 
with superimposed the  intermittent slope $\zeta_1=0.39$.}
\label{fig:3}
\end{figure}
%%%%%%%%%%%%%%%%%%%%%%%%

In Figure~\ref{fig:3} we show  $\Sigma_1(\delta v)$ 
evaluated  on a string of high-Reynolds number experimental data as a function
of the available range of velocity thresholds $\delta v$. 
This data set has been  measured
in a wind tunnel  at $Re_{\lambda} \sim 2000$.  
One can see that the scaling is very poor.
On the other hand, (inset of Figure~\ref{fig:3}),  the 
scaling behavior of the direct structure functions 
$<\!|\delta v (\tau)|\!> \sim \tau^{\zeta_1}$ 
is quite clear in a wide range of scales. 
This is a clear evidence of  IDR's contamination  into the 
whole range of available velocity values for the Inverse-SF cases.

Let us now go back to  the
statistical properties of the IDR. 
In order to study this question we have smoothed the 
stochastic synthetic field, $v(t)$ (see Appendix)  by performing a 
running-time average
over a time-window, $\delta T$. 
Then we compare Inverse-SF obtained for different Reynolds numbers, 
i.e. for different dissipative cut-off: $Re \sim \delta T^{-4/3}$.\\
The expression (\ref{II.10}) predicts the possibility
to obtain a data collapse of all curves with 
different Reynolds numbers by rescaling  the Inverse-SF as
follows \cite{FV91,JPV91}: 
\begin{equation}
- {\ln (\Sigma_p(\delta v))} / {\ln (\delta T / \delta T_0)} \;\;vs. \;\; 
-{\ln(\delta v/U)} / {\ln (\delta T / \delta T_0)}\,,
\label{II.11}
\end{equation}
where $U$ and  $\delta T_0$ are adjustable dimensional parameters.

%%%%%%%%%%%%%%%%%%%%%%%%
\begin{figure}[!ht]
\centering
\includegraphics[width=10 cm]{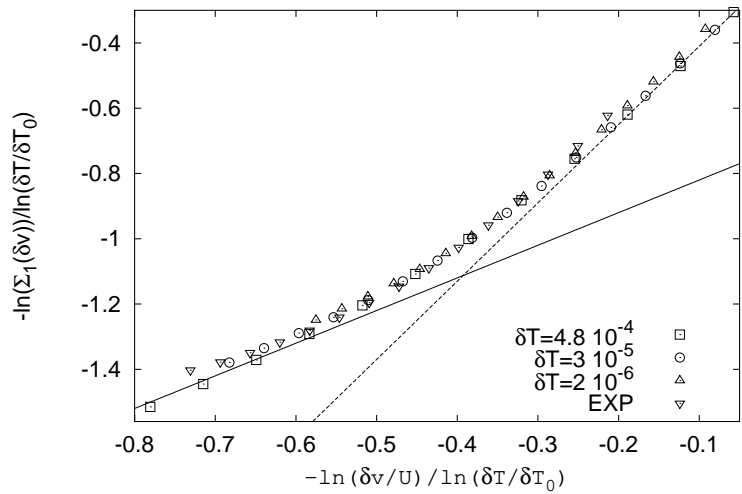}
\caption{Data collapse of 
the Inverse-SF, $\Sigma_1(\delta v)$,
 obtained by the rescaling (\ref{II.11}) for the
smoothed synthetic signals (with time windows: 
$\delta T=4.8\cdot 10^{-4}\,,\;3\cdot 10^{-5}\,,\; 2\cdot 10^{-6}$) 
and the experimental
data ($EXP$). The two straight lines have the dissipative (solid line)
and the  inertial range (dashed) slope.}
\label{fig:4}
\end{figure}
%%%%%%%%%%%%%%%%%%%%%%%%

Figure \ref{fig:4}  shows the rescaling (\ref{II.11}) of the Inverse-SF, 
$\Sigma_1(\delta v)$,
both for the synthetic  field at different Reynolds numbers and 
for the experimental signals.
As it is possible to see, the data-collapse is very good. 
This is a clear evidence that the poor scaling range observed in 
Figure \ref{fig:4}
for the experimental signal can be explained as the signature of the IDR. 

%%%%%%%%%%%%%%%%%%%%%%%%%%%%%%%%%%%%%%%%%%%%%%%%%%%%%%%%%%%%%%%%%%%%%%%%
\section{The relation between Eulerian and Lagrangian statistics}
\label{sec:4}

A problem of great interest concerns the study of the spatial and 
temporal structure of the so-called passive fields, indicating by 
this term quantities transported by the flow without affecting the
velocity field. The paradigmatic equation for the evolution of a passive
scalar field $\theta ({\bf x},t)$ advected by a velocity field 
${\bf v}({\bf x},t)$ is \cite{ShraimanSiggia00}
\begin{equation}
\partial_t\theta +\nabla \cdot ({\bf v}\,\theta) =  \chi\,\nabla^2\theta
\label{eq:4.1.1}
\end{equation}
where $\chi$ is the molecular diffusion coefficient.

The problem (\ref{eq:4.1.1}) can be studied through two equivalent
approaches, both due to Euler \cite{Lamb45}. 
The first, referred to as ``Eulerian'', deals at any time with 
the field $\theta$ in the space domain covered by the fluid; the
second considers the time evolution of trajectories of each fluid
particle and is called ``Lagrangian''.

The motion of a fluid particle is determined by the differential
equation
\begin{equation}
\frac{d{\bf x}}{d t}= {\bf v}({\bf x},t)
\label{eq:4.1.2}
\end{equation}
which also describes the motion of test particles, for example a
powder embedded in the fluid, provided that the particles are neutral and
small enough not to perturb the velocity field, although large enough not 
to perform a Brownian motion. Particles of this type are commonly used 
for flow visualization in fluid mechanics experiments \cite{T88}.
We remark that the complete equation for the motion of a material particle 
in a fluid when density and volume effects are taken into account can be 
rather complicated \cite{MR83,BBCLT06a}.

The Lagrangian equation of motion (\ref{eq:4.1.2}) formally 
represents a dynamical system in the phase space of physical
coordinates. By very general considerations, it is now well 
established that even in regular velocity field
the motion of fluid particles can be very irregular \cite{H66,A84}.
In this case initially nearby trajectories diverge exponentially 
and one speaks of {\it Lagrangian chaos} or {\it chaotic advection}.
In general, chaotic behaviors can arise in two-dimensional flow only 
for time dependent velocity fields, while it can
be present even for stationary velocity fields in three dimensions.

If $\chi=0$, it is easy to realize that (\ref{eq:4.1.1}) is equivalent to 
(\ref{eq:4.1.2}). Indeed, we can write
\begin{equation}
\theta({\bf x},t)= \theta({T}^{-t} {\bf x},0)
\label{eq:4.1.3}
\end{equation}
where ${T}$ is the formal evolution operator of (\ref{eq:4.1.2}):
${\bf x}(t)= {T}^t{\bf x}(0)$.

Taking into account the molecular diffusion $\chi$, 
(\ref{eq:4.1.1}) is the Fokker-Planck equation of the Langevin 
 equation \cite{Chandrasekhar43} 
\begin{equation}
{d{\bf x} \over d t}= {\bf v}({\bf x}, t) + {\eta}(t)
\label{eq:4.1.4}
\end{equation}
where ${\eta}$ is a Gaussian process with zero mean and variance
\begin{equation}
\left\langle{\eta_i(t)\, \eta_j(t')}\right\rangle= 
 2\chi \delta_{ij}\,\delta(t-t').
\label{eq:4.1.5}
\end{equation}

The dynamical system (\ref{eq:4.1.2}) becomes conservative in the
phase space in the case of an incompressible velocity field for which
\begin{equation}
\nabla\cdot{\bf v}= 0
\label{eq:4.1.6}
\end{equation}
In two dimensions, ${\bf x}=(x_1,x_2)$, the constraint (\ref{eq:4.1.6}) 
is automatically satisfied by introducing the stream function 
$\psi({\bf x},t)$
\begin{equation}
v_1= {\partial \psi\over\partial x_2}, \qquad
v_2= -\,{\partial \psi\over\partial x_1}
\label{eq:4.1.7}
\end{equation}
and the evolution equation becomes
\begin{equation}
{ {d x_1} \over {d t} }= {\partial \psi\over\partial x_2}, \qquad
     { {d x_2} \over {d t} }= -\,{\partial \psi\over\partial x_1}.
\label{eq:4.1.8}
\end{equation}
i.e. formally a Hamiltonian system with the Hamiltonian given by 
the stream function $\psi$.
 
The presence of Lagrangian chaos in regular flows is a remarkable
example of the fact that, in general, it is very difficult to
relate Lagrangian and Eulerian statistics. For example, from
very complicated trajectories of buoys one cannot infer
the time-dependent circulation of the sea.
In the following Sections we will see that in the case of 
fully developed turbulence, the disordered nature of the
flow makes this connection partially possible 
at a statistical level.

The equation of motion (\ref{eq:4.1.2}) shows that the 
trajectory of a single particle is not Galilean invariant,
i.e. invariant with respect to the addition of a mean velocity.
The most general Galilean invariant statistics, which is 
ruled by small scale velocity fluctuation, is given
by multi-particle, multi-time correlations for which we 
could expect universal features. In the following we will consider
separately the two most studied statistics: single-particle
two-time velocity differences and two-particle single-time relative
dispersion.

%%%%%%%%%%%%%%%%%%%%%%%%%%%%%%%%%%%%%%%%%%%%%%%%%%%%%%%%%%%%%%
%%%%%%%%%%%%%%%%%%%%%%%%%%%%%%%%%%%%%%%%%%%%%%%%%%%%%%%%%%%%%%
\subsection{Single particle statistics: multifractal description 
of Lagrangian velocity differences}
\label{sec:4.2}
The simplest Galilean invariant Lagrangian quantity is the
single particle velocity increment 
$\delta {\bf v}(t)={\bf v}(t_0+t)-{\bf v}(t_0)$, where 
${\bf v}(t)={\bf v}({\bf x}(t),t)$ denotes the Lagrangian 
velocity of the particle at ${\bf x}(t)$ and the independence on
$t_0$ is a consequence of the stationarity of the flow.
Dimensional analysis in fully developed turbulence predicts
\cite{MY75,TL72}
\begin{equation}
\langle \delta v_{i}(t) \delta v_{j}(t) \rangle =
C_{0} \bar{\varepsilon} t \delta_{ij}
\label{eq:4.2.1}
\end{equation}
where $\bar{\varepsilon}$ is the mean energy dissipation and $C_{0}$ is
a numerical constant. The remarkable coincidence that the
variance of $\delta {\bf v}(t)$ grows linearly with time
is the physical basis on which stochastic models of particle
dispersion are based.
It is important to recall that the
``diffusive'' nature of (\ref{eq:4.2.1}) is purely incidental:
it is a direct consequence of Kolmogorov scaling in the inertial
range of turbulence and is not directly related to a diffusive
process (i.e. there is no decorrelation justifying the applicability 
of central limit theorem).

Let us recall briefly the argument leading
to the scaling in (\ref{eq:4.2.1}). We can think at the velocity
$v(t)$ advecting the Lagrangian trajectory as the superposition
of the different velocity contributions coming from turbulent eddies
(which also move with the same velocity of the Lagrangian trajectory).
After a time $t$ the components associated to the smaller
(and faster) eddies, below a certain scale $\ell$ are decorrelated
and thus at the leading order one has
$\delta v(t) \simeq \delta v(\ell)$. Within Kolmogorov scaling,
the velocity fluctuation at scale $\ell$ is given by
$\delta v(\ell) \sim V_0 (\ell/L)^{1/3}$ where $V_0$
represents the typical velocity at the largest scale $L$.
The correlation time of $\delta v(\ell)$ scales as
$\tau(\ell) \sim \tau_0 (\ell/L)^{2/3}$ and thus one obtains
the scaling in (\ref{eq:4.2.1}) with $\varepsilon=V_0^2/\tau_0$.

Equation (\ref{eq:4.2.1}) can be generalized to higher order moments
with the introduction of a set of temporal scaling exponents $\xi_p$
\begin{equation}
\langle \delta v(t)^{p} \rangle \sim V_0^p (t/\tau_0)^{\xi_p}
\label{eq:4.2.2}
\end{equation}
The dimensional estimation sketched above gives the prediction $\xi_p=p/2$
but one may expect corrections to the dimensional scaling in the presence of
intermittency.

A generalization of the above results which takes into account 
intermittency corrections can be easily developed by using the
multifractal model \cite{Borgas93,BDM02}. 
The dimensional argument is repeated for the local
scaling exponent $h$, giving $\delta v(t) \sim V_0 (t/\tau_0)^{h/(1-h)}$.
Integrating over the $h$ distribution one ends with
\begin{equation}
\langle \delta v(t)^{p} \rangle \sim V_0^p
\int dh \left({t \over \tau_0} \right)^{[ph-D(h)+3]/(1-h)} \, .
\label{eq:4.2.3}
\end{equation}
where $D(h)$ is the Eulerian fractal dimension (i.e. related to the
Eulerian structure function scaling exponents by 
$\zeta(q)=\min_{h}[qh-D(h)+3]$).
In the limit $t/\tau_0 \to 0$, the integral can be estimated
by a steepest descent argument giving the prediction
\begin{equation}
\xi_p=\min_{h}\left[{ph-D(h)+3 \over 1-h} \right]
\label{eq:4.2.4}
\end{equation}
The standard inequality in the multifractal model
$D(h)\le 3h+2$ implies for (\ref{eq:4.2.4})
that even in the presence of intermittency $\xi_2=1$. 
Physically, this is a consequence of the fact that energy dissipation 
is raised to the first power, in (\ref{eq:4.2.1}).

Experimental results \cite{MMMP01} have shown that even at 
large Reynolds number the scaling (\ref{eq:4.2.1}) is not
clearly observed. Therefore the dimensionless constant $C_0$
is known with large uncertainty, if compared with the Kolmogorov
constant. 

\begin{figure}[ht]
\includegraphics[width=10cm]{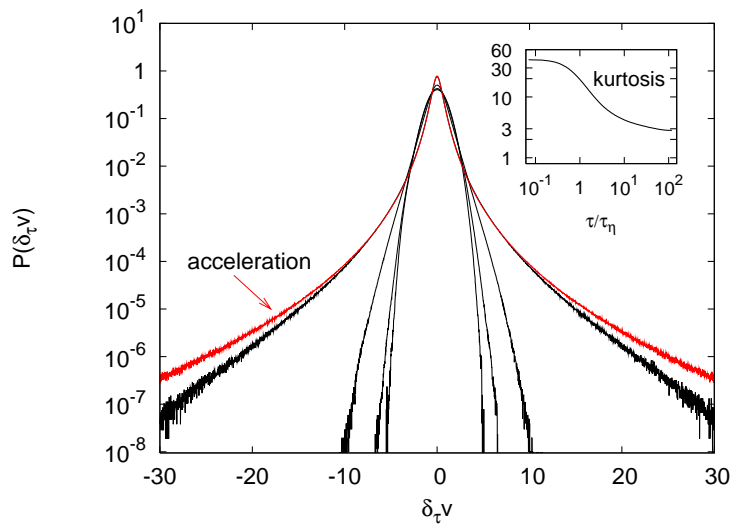}
\caption{Probability density functions of velocity increments 
for a numerical simulation at $R_{\lambda}=284$. 
Curves refer to time increments $t = (97, 24, 6, 0.7) \tau_{D}$
from inside to outside (and to the acceleration). 
In the inset kurtosis $\langle \delta v(t)^4 \rangle /
\langle \delta v(t)^2 \rangle^2$ as a function of time delay is
shown. Figure from \cite{BBCLT06}.}
\label{fig:4.2.2}
\end{figure}

Intermittency in Lagrangian velocity differences is evident
by looking at the pdf of $\delta v(t)$ at different time lags,
as shown in Fig.~\ref{fig:4.2.2}. For large time delays the pdf are 
close to Gaussian while decreasing $t$ they develop larger and
larger tails, implying the breakdown of self-similarity.
%%%%%%%%%%%%%%%%%%%%%%%%%%%%%%%%%%%%%%%%%%%%%%%%%%%%%%%%%%%%
%%%%%%%%%%%%%%%%%%%%%%%%%%%%%%%%%%%%%%%%%%%%%%%%%%%%%%%%%%%%

\begin{figure}[ht]
\includegraphics[width=10cm]{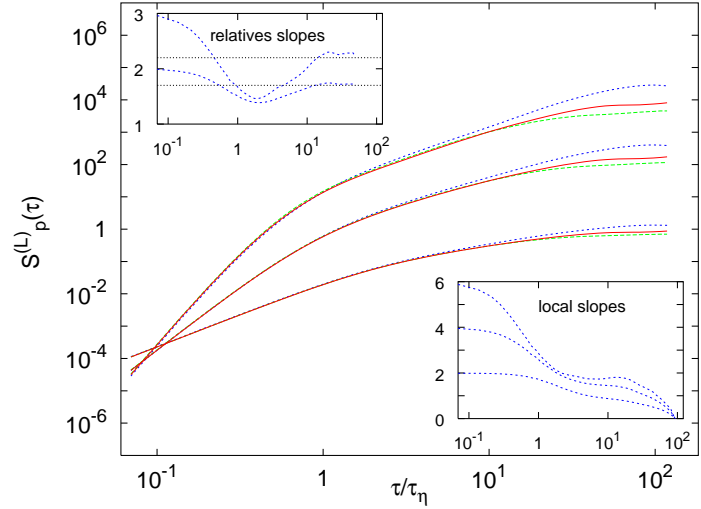}
\caption{Log-log plot of Lagrangian structure functions of order 
$2$, $4$ and $6$ (bottom to top) for the three components of the 
velocity obtained from DNS at $R_{\lambda}=284$. Small differences
between the components are observed at large delays as a consequence
of small anisotropies at large scales. In the insets the local
slopes (bottom right) and relative slopes (top left) are shown 
for the $x$-component. Relative slopes are defined as
$d S_p(t)/d S_2(t)$ for $p=4,6$. The horizontal lines represent 
the multifractal prediction. Figure from \cite{BBCLT06}.}
\label{fig:4.2.3}
\end{figure}

Higher order Lagrangian structure functions are shown in 
Fig.~\ref{fig:4.2.3} for a set of direct numerical simulations
at $R_{\lambda}=284$ \cite{BBCLT05,BBCLT06}. Despite the apparent scaling
observed in the log-log plot, the computation of local slopes does
not give a definite value of scaling exponents. Assuming 
$\xi_2=1$ as predicted by (\ref{eq:4.2.4}),
one can measure the {\it relative} scaling exponent $\xi_p/\xi_2$
by using the so-called  extended self-similarity procedure \cite{BCTBMS93}.
As shown by the inset of Fig.~\ref{fig:4.2.3}, we
observe a well defined scaling in the range of separations
$10\tau_{D} \le \tau \le 50\tau_{D}$.
The values of the relative exponents estimated with this method,
$\xi_4/\xi_2=1.7\pm0.05$,
$\xi_5/\xi_2=2.0\pm0.05$,
$\xi_6/\xi_2=2.2 \pm0.07$, are in good agreement with those
predicted by the multifractal model (\ref{eq:4.2.4}).

%%%%%%%%%%%%%%%%%%%%%%%%%%%%%%%%%%%%%%%%%%%%%%%%%%%%%%%%%%%%
%%%%%%%%%%%%%%%%%%%%%%%%%%%%%%%%%%%%%%%%%%%%%%%%%%%%%%%%%%%%
The multifractal prediction (\ref{eq:4.2.4}) has been checked 
also in simplified Lagrangian model based on the shell model
of turbulence \cite{BDM02}. 

%%%%%%%%%%%%%%%%%%%%%%%%%%%%%%%%%%%%%%%%%%%%%%%%%%%%%%%%%%%%%%
%%%%%%%%%%%%%%%%%%%%%%%%%%%%%%%%%%%%%%%%%%%%%%%%%%%%%%%%%%%%%%
\section{Lagrangian acceleration statistics}
\label{sec:4.3}
Acceleration in fully developed turbulence is an extremely intermittent
quantity which display fluctuations up to $80$ times its root
mean square \cite{LVCAB01}. These extreme events generate
very large tails in the pdf of acceleration which are therefore expected
to be very far from Gaussian. 

We remark that even within non-intermittent Kolmogorov scaling,
acceleration pdf is expected to be non-Gaussian. Indeed acceleration
can be estimated from velocity fluctuations at the Kolmogorov scale
as
\begin{equation}
a = {\delta v(\tau_{D}) \over \tau_{D}}
\label{eq:4.3.1}
\end{equation}
where $\tau_{D}=\eta/\delta v(\eta)$ and the Kolmogorov
scale $\eta$ is given by the condition $\eta \delta v(\eta)/\nu=1$.
By assuming the scaling $\delta v(\ell) \simeq V_0 (\ell/L)^h$
(with $h=1/3$ for Kolmogorov scaling) one obtains
\begin{equation}
{\eta \over L} \sim \left({V_0 L \over \nu}\right)^{-{1 \over 1+h}}
\label{eq:4.3.2}
\end{equation}
and therefore
\begin{equation}
a = {V_0^2 \over L} \left({V_0 L \over \nu}\right)^{1-2h \over 1+h}
\label{eq:4.3.3}
\end{equation}
Assuming a Gaussian distribution for large scale velocity 
fluctuations $V_0$ (which is, as already observed, consistent with many experimental and
numerical observations), and taking $h=1/3$, one obtains for the 
pdf of $a$ a stretched
exponential tail $p(a) \sim \exp(-C a^{8/9})$.

In the presence of intermittency the above argument has to be 
modified by taking into
 account the fluctuations of scaling exponent.
In the recent years, several models have been proposed for describing 
turbulent acceleration statistics, on the basis of different
physical ingredients. In the following we want to show that
the multifractal model of turbulence, when extended to describe
fluctuation at the dissipative scale, is able to predict the pdf 
of acceleration observed in simulations and experiments with high
accuracy \cite{BBCDLT04}. Moreover, as in the case of Lagrangian
structure functions, the model does not require the introduction of
new parameters, a part the set of Eulerian scaling exponents. In
this sense, multifractal model become a {\it predictive} model for
Lagrangian statistics.

The introduction of intermittency in the above argument is simply
obtained by weighting (\ref{eq:4.3.3}) with both the distribution 
of $V_0$ (still assumed Gaussian, as intermittency is not expected
to affect large scale statistics) and the distribution of scaling
exponent $h$ which can be rewritten as
\begin{equation}
p(h) \sim \left({\eta \over L} \right)^{3-D(h)} \sim 
\left({V_0 L \over \nu}\right)^{D(h)-3 \over 1+h}
\label{eq:4.3.4}
\end{equation}
The final prediction, when written for the dimensionless acceleration
$\tilde{a}=a/\langle a^2 \rangle^{1/2}$, becomes \cite{BBCDLT04}
\begin{equation}
p(\tilde{a}) \sim \int_{h} \tilde{a}^{[h-5+D(h)]/3}
R_{\lambda}^{y(h)} \exp \left(-{1 \over 2} \tilde{a}^{2(1+h)/3}
R_{\lambda}^{z(h)}\right) dh
\label{eq:4.3.5}
\end{equation}
where $y(h) = \chi (h-5+D(h))/6 + 2(2D(h)+2h-7)/3$ and
$z(h) = \chi (1+h)/3 + 4(2h-1)/3$. The coefficient $\chi$
is the scaling exponent for the Reynolds dependence of the 
acceleration variance,
$\langle a^2 \rangle \sim R_{\lambda}^{\chi}$, given by
$\chi = \sup_h \left( 2 (D(h)-4h-1)/(1+h) \right)$.
For the non-intermittent Kolmogorov scaling ($h=1/3$ and $D(1/3)=3$)
one obtains $\chi=1$ and (\ref{eq:4.3.5}) recovers the stretched 
exponential prediction discussed above.

We note that (\ref{eq:4.3.5}) may show an unphysical divergence for
$a \to 0$ for many multifractal models of $D(h)$ at small $h$.
This is not a real problem 
for two reasons. First, the multifractal formalism cannot be
extended to very small velocity and acceleration increments because it is
based on arguments valid only to within a constant of order one.
Thus, it is not suited for predicting precise
functional forms for the core of the pdf. Second, small values of
$h$ correspond to very intense velocity
fluctuations which have never been accurately tested in experiments or by
DNS. The precise functional form of $D(h)$
for those values  of $h$ is therefore unknown. 

\begin{figure}[ht]
\includegraphics[width=10cm]{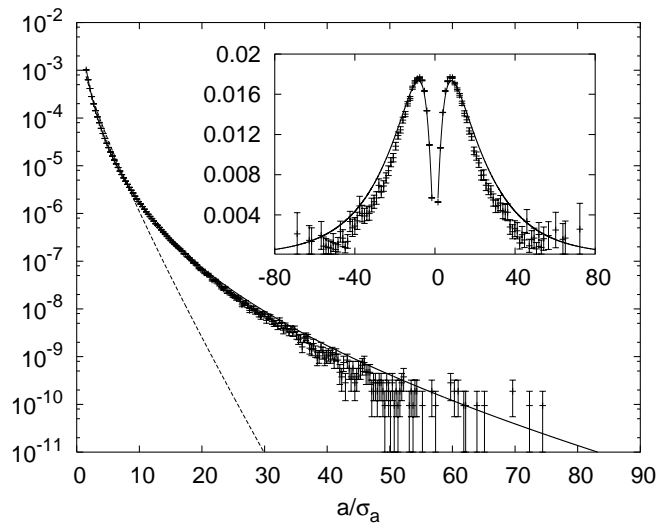}
\caption{Log-linear plot the acceleration pdf. Points
are the DNS data, the solid line is the multifractal prediction and
the dashed line is the K41 prediction.
The DNS statistics were calculated along the trajectories of two 
million particles amounting to $3.6 \times 10^9$ events in total. 
The statistical
uncertainty in the pdf was quantified by assuming that fluctuations grow
like the square root of the number of events.
Inset: ${\tilde a}^4 p({\tilde a})$ for the DNS data (crosses) and the
multifractal prediction.}
\label{fig:4.3.1}
\end{figure}

In Fig.~(\ref{fig:4.3.1}) we compare the acceleration pdf computed from
the DNS data at $R_{\lambda}=280$ with the multifractal prediction 
(\ref{eq:4.3.5}) using for $D(h)$ an empirical model which fits well
the Eulerian scaling exponents \cite{SheLeveque94}.
The large number of Lagrangian particles used in the DNS 
(see \cite{BBCLT05} for details) allows us to detect events up to $80$
$\sigma_a$. The accuracy of the statistics is improved by averaging
over the total duration of the simulation and all directions since the
flow is stationary and isotropic at small scales.  Also shown in
Fig.~(\ref{fig:4.3.1}) is the non-intermittent prediction
$p(\tilde{a}) \simeq \tilde{a}^{-5/9} R_{\lambda}^{-1/2}
\exp \left( -\tilde{a}^{8/9}/2 \right)$.
As is evident from the figure, the multifractal prediction
captures the shape of the acceleration pdf much better than the K41
prediction. What is remarkable is that (\ref{eq:4.3.5})  agrees
with the DNS data well into the tails of the distribution -- from the
order of one standard deviation $\sigma_a$ up to order $70 \sigma_a$.
We emphasize that the only free parameter in the multifractal formulation 
of $p(\tilde{a})$ is the minimum value of the acceleration, ${\tilde
a}_{\min}$, here taken to be $1.5$. 
In the inset of Fig.~(\ref{fig:4.3.1}) we
make a more stringent test of the multifractal prediction
(\ref{eq:4.3.5}) by plotting  ${\tilde a}^4 p({\tilde a})$ and which is
seen to agree well with the DNS data.

%%%%%%%%%%%%%%%%%%%%%%%%%%%%%%%%%%%%%%%%%%%%%%%%%%%%%%%%%%%%
%%%%%%%%%%%%%%%%%%%%%%%%%%%%%%%%%%%%%%%%%%%%%%%%%%%%%%%%%%%%

%%%%%%%%%%%%%%%%%%%%%%%%%%%%%%%%%%%%%%%%%%%%%%%%%%%%%%%%%%%%%%
%%%%%%%%%%%%%%%%%%%%%%%%%%%%%%%%%%%%%%%%%%%%%%%%%%%%%%%%%%%%%%
\section{Relative dispersion in turbulence}
\label{sec:4.4}
Relative dispersion of two particles is historically the first issue
quantitatively addressed in the study of fully developed turbulence.
This was done by Richardson, in a pioneering work on the properties of
dispersion in the atmosphere in 1926 \cite{Richardson26}, 
and then reconsidered by Batchelor \cite{Batchelor52}, among others, 
in the light of Kolmogorov 1941 theory \cite{F95}.

Richardson's description of relative dispersion is based on
a diffusion equation for the probability density function 
$p({\bf r},t)$ where ${\bf r}(t)={\bf x}_{2}(t)-{\bf x}_{1}(t)$ is the
separation of two trajectories generated by (\ref{eq:4.1.2}). 
In the isotropic case the diffusion equation can be written as
\begin{equation}
{\frac{\partial p({\bf r},t)}{\partial t}}={\frac{1}{r^{2}}}{\frac{\partial
}{\partial r}}r^{2}K(r){\frac{\partial p({\bf r},t)}{\partial r}},
\label{eq:4.4.1}
\end{equation}
where the turbulent eddy diffusivity was empirically established by
Richardson to follow the ``four-thirds law'':
$K(r)=k_{0}\varepsilon^{1/3}r^{4/3}$ in which $k_0$ is a 
dimensionless constant.
The scale dependence of diffusivity is at the origin of the
accelerated nature of turbulent dispersion: particle relative
velocity grows with the separation. Richardson
empirical formula is a simple consequence of Kolmogorov scaling
in turbulence, as first recognized by Obukhov \cite{Obukhov41}.

The solution of (\ref{eq:4.4.1}) for $\delta$-distributed
initial condition has the well known stretched exponential form
\begin{equation}
p({\bf r},t)={\frac{A}{(k_{0}\varepsilon^{1/3}t)^{9/2}}}
\exp \left( -{\frac{9r^{2/3}}{4k_{0}\varepsilon ^{1/3}t}}\right)
\label{eq:4.4.2}
\end{equation}
where $A=2187/2240 \pi^{3/2}$ is a normalizing factor.
Of course, the assumption the relative dispersion can be described
by a self-similar process as (\ref{eq:4.4.1}) rules out the possibility 
of intermittency and therefore the scaling exponents of the moments of 
relative separation 
\begin{equation}
\langle r^{2n}(t) \rangle = C_{2n} \varepsilon^n t^{\alpha_n}
\label{eq:4.4.3}
\end{equation}
have the values $\alpha_{n}=3n$, as follows from dimensional analysis.
All the dimensionless coefficients $C_{2n}$ are in this case given in terms of
$k_0$ and a single number, such as the so-called Richardson constant
$C_2$, is sufficient to parameterize turbulent dispersion.

The hypothesis of self-similarity is reasonable in the presence of a 
self-affine Eulerian velocity field, such as in the case of two-dimensional
inverse cascade where the dimensional exponents $\alpha_{2n}=3n/2$
have indeed been found \cite{BS_POF02}. An analysis of Lagrangian 
trajectories generated by a kinematic model with synthetic velocity
field \cite{BCCV_PRE99} has shown that Lagrangian self-similarity
is broken in the presence of Eulerian intermittency. In this case
it is possible to extend the dimensional prediction for the scaling
exponents $\alpha_n$ by means of the multifractal model of turbulence.

From the definition of relative separation
\begin{equation}
{d \over dt} \langle r^{p}(t) \rangle = \langle r^{p-1} \delta v(r) \rangle
\label{eq:4.4.4}
\end{equation}
where $\delta v(r)$ is the velocity increments between the two trajectories. 
Using the multifractal representation (\ref{I.5}) we can write
\begin{equation}
{d \over dt} \langle r^{p}(t) \rangle \sim \int dh \,
r^{p-1+h+3-D(h)}
\label{eq:4.4.5}
\end{equation}
The time needed for the pair separation to reach the scale $r$
is dominated by the largest time in the process, associated to 
the scale $r$ and therefore given by $t \sim r^{1-h}$. This leads
to
\begin{equation}
{d \over dt} \langle r^{p}(t) \rangle \sim \int dh \,
t^{[p+2+h-D(h)]/(1-h)}
\label{eq:4.4.6}
\end{equation}
The integral is evaluated by saddle point method and gives
the final result $\langle r^{p}(t) \rangle \sim t^{\alpha_p}$
with scaling exponents
\begin{equation}
\alpha_p = \inf_{h}\left[ {p+3-D(h) \over 1-h} \right] 
\label{eq:4.4.7}
\end{equation}
From the standard inequality of the multifractal formalism
(\ref{I.8}) one obtains that even in the presence of intermittency
$\alpha_2=3$. As in the case of single particle dispersion (\ref{eq:4.2.4})
also here this is a consequence of the presence on the first
power of $\varepsilon$ in (\ref{eq:4.4.3}) for $n=1$. 

The scaling exponents $\alpha_p$ satisfy the inequality
$\alpha_p/p < 3/2 $ for $p>2$. This amounts to say
that, as time goes on, the right tail of the particle pair separation
probability distribution function becomes narrower and narrower.
In other words, due to the Eulerian intermittency particle pairs are more
likely to stay close to each other than to experience a large
separation.
%%%%%%%%%%%%%%%%%%%%%%%%%%%%%%%%%%%%%%%%%%%%%%%%%%%%%%%%%%%%
\begin{figure}[ht]
\includegraphics[width=10cm]{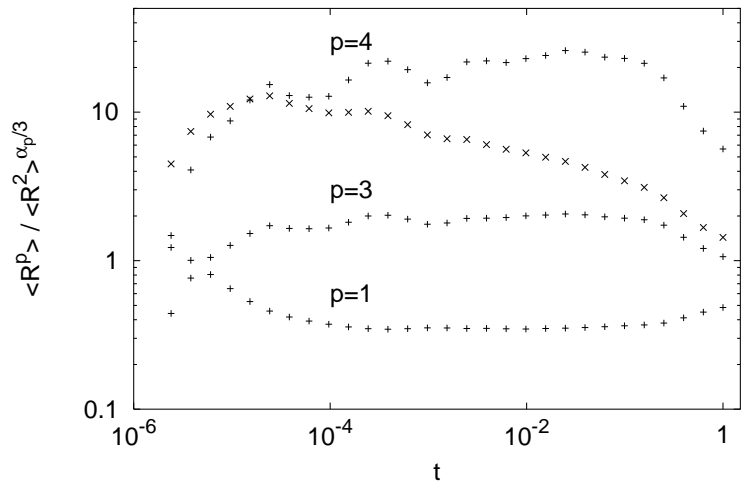}
\caption{Relative dispersion $\langle r^p(t) \rangle$ rescaled with
$\langle r^2(t) \rangle^{\alpha_p/3}$ for $p=1,3,4$ ($+$)
for Lagrangian simulations in a synthetic velocity field
corresponding to $R_{\lambda}\approx 10^{6}$.
The almost
constant plateau indicates a relative scaling in agreement with
prediction (\ref{eq:4.4.7}). For comparison we also plot
$\langle r^4(t) \rangle$ rescaled with the non intermittent prediction
$\langle r^2(t) \rangle^2$ ($\times$) clearly indicating a
deviation from normal scaling. Figure from \cite{BCCV_PRE99}.}
\label{fig:4.4.1}
\end{figure}
%%%%%%%%%%%%%%%%%%%%%%%%%%%%%%%%%%%%%%%%%%%%%%%%%%%%%%%%%%%%

The multifractal prediction (\ref{eq:4.4.7}) has been checked in 
synthetic model of fully developed turbulence \cite{BCCV_PRE99} 
where the equivalent Reynolds number is very large. In the case
of numerical or experimental data, finite Reynolds effects make 
very difficult to measure the corrections to dimensional exponents.
We remark that finite Reynolds effects are more important in 
Lagrangian dispersion than in Eulerian statistics: as a consequence
of the accelerate nature of relative motion a large fraction of 
pairs exits the inertial range after a short time.

To overcome these difficulties in Lagrangian statistics, an alternative
approach based on exit time statistics has been proposed 
for Lagrangian dispersion \cite{ABCCV97,BCCV_PRE99}. 
In close analogy with the exit time approach described in 
Section~\ref{sec:2.3}, one computes the
doubling times $T_{\rho }(R_{n})$ for a pair separation 
to grow from threshold $R_{n}$ to the next one $R_{n+1}$.
Averages are then performed over many particle
pairs. The outstanding advantage
of averaging at fixed scale separation, 
as opposed to averaging at a fixed time, is that crossover effects are removed
since all sampled particle pairs
belong to the same scales. \\
Neglecting intermittency, the doubling time analysis can be used for
a precise estimation of the Richardson constant $C_{2}$. From the
first-passage problem for the Richardson model (\ref{eq:4.4.1})
one has \cite{BS_PRL02}:
\begin{equation}
\langle T_{\rho }(R)\rangle ={\frac{\rho ^{2/3}-1}{2k_{0}
\varepsilon^{1/3}\rho ^{2/3}}}R^{2/3}
\label{eq:4.4.8}
\end{equation}
from which one obtains
\begin{equation}
C_{2}={\frac{143}{81}}{\frac{(\rho ^{2/3}-1)^{3}}{\rho ^{2}}}
{\frac{R^{2}}{\varepsilon \langle T_{\rho }\rangle ^{3}}}.
\label{eq:4.4.9}
\end{equation}
By using this expression it is possible to estimate from DNS data
at moderate Reynolds $C_2 = 0.50 \pm 0.05$ \cite{BS_PRL02,BBCDLT_POF05}
which is in agreement with the experimental determination \cite{OM_JFM00}.

Intermittency effects are evident in higher order statistics of 
doubling times. In particular, one expects for the moments of inverse 
doubling times, $\langle \left( 1/T_{\rho }(R)\right)^{p}\rangle$
a power-law behavior
\begin{equation}
\langle \left( {\frac{1}{T_{\rho }(R)}}\right)^{p}\rangle
\sim R^{\beta_{p}}
\label{eq:4.4.10}
\end{equation}
with exponents $\beta _{p}$ connected to the exponents $\alpha _{n}$
\cite{BCCV_PRE99}.
Negative moments of doubling time are dominated by pairs which
separate fast; this corresponds to positive moments of relative separation.
By using the simple dimensional estimate $T(R)\sim R/\delta v(R)$
one has the prediction
\begin{equation}
\beta _{p}=\zeta _{p}-p ,
\label{eq:4.4.11}
\end{equation}
where $\zeta_{p}$ are the scaling exponents of the Eulerian structure
functions (\ref{eq:1.1.5}). 

%%%%%%%%%%%%%%%%%%%%%%%%%%%%%%%%%%%%%%%%%%%%%%%%%%%%%%%%%%%%
\begin{figure}[ht]
\includegraphics[width=10cm]{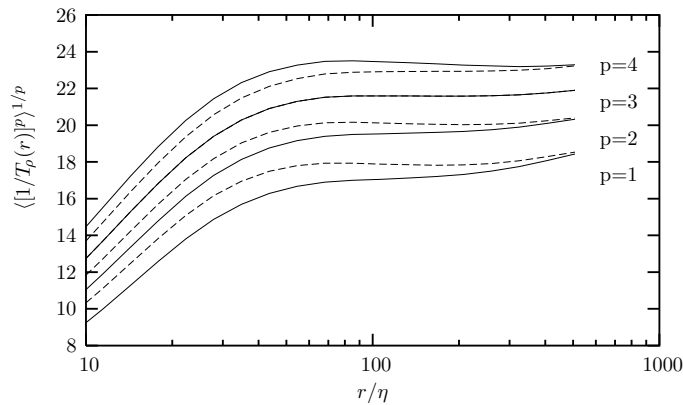}
\caption{The inverse exit time moments, $\langle [1/T_\rho(r)]^p
\rangle^{1/p}$, for $p=1,\ldots,4$ compensated with the Kolmogorov
scalings (solid lines) and the multifractal predictions (dashed lines).
Results from DNS at $R_{\lambda}=284$, initial particle separation
$r_0 = 1.2\eta$ and ratio between exit times $\rho=1.25$.
Figure from \cite{BBCDLT_POF05}.}
\label{fig:4.4.2}
\end{figure}
%%%%%%%%%%%%%%%%%%%%%%%%%%%%%%%%%%%%%%%%%%%%%%%%%%%%%%%%%%%%

The multifractal prediction (\ref{eq:4.4.11}) is found to be 
consistent with numerical data at moderate Reynolds number. More 
important, as shown in Figure~\ref{fig:4.4.2}, exit-time statistics
is sufficiently accurate for discriminating between intermittent
and dimensional scaling in Lagrangian statistics.

%%%%%%%%%%%%%%%%%%%%%%%%%%%%%%%%%%%%%%%%%%%%%%%%%%%%%%%%%%%%%%
%%%%%%%%%%%%%%%%%%%%%%%%%%%%%%%%%%%%%%%%%%%%%%%%%%%%%%%%%%%%%%
\section{Dispersion in two-dimensional convection: 
multifractal analysis of more-than-smooth signals}
\label{sec:4.5}
Thermal convection in two-dimensions provides an example of 
Bolgiano-Obukhov scaling of turbulent fluctuations. 
Without entering in the details, we recall that within
Boussinesq approximation, Bolgiano-Obukhov argument assumes
a local balance between buoyancy force and inertial term
\cite{Siggia94}. In the case of two-dimensional turbulence,
in the presence of a mean temperature gradient, Bolgiano-Obukhov
scaling is expected to emerge in the inverse cascade of energy
with velocity fluctuations given by the scaling law \cite{Chertkov03}:
\begin{equation}
\delta v(r) \propto {\varepsilon}_T^{1/5} (\beta g)^{2/5} r^{3/5} \, ,
\label{eq:4.5.1}
\end{equation}
where ${\varepsilon}_T$
is the (constant) flux of temperature fluctuations, $\beta$ is the 
thermal expansion coefficient and $g$ is the gravity acceleration.
The prediction (\ref{eq:4.5.1}) has been checked in both laboratory
experiments \cite{ZWX05} and in high resolution direct numerical
simulations \cite{CMV01} which have also shown the absence of 
intermittency corrections (which is a common feature of two-dimensional 
inverse cascades).

We now consider the increments of velocity for Lagrangian 
tracers transported by Bolgiano turbulence. By extending the
dimensional argument of Section~\ref{sec:4.2} to the general case of 
velocity scaling exponent $h$ one obtains \cite{BBM_POF07}
\begin{equation}
\delta v(t) \propto V (t/\tau_0)^{h/(1-h)}
\label{eq:4.5.1b}
\end{equation}
At variance with Navier-Stokes turbulence, from (\ref{eq:4.5.1}) $h=3/5$ and
therefore $q=h/(1-h)=3/2>1$, i.e. velocity increments in the 
inertial range are smoother than $C_1$ signals, the latter denoting the class
of differentiable signals. This implies that
Lagrangian structure functions (\ref{eq:4.2.2}) are
dominated by non-local contributions from the large scale $L$ 
which scale as
\begin{equation}
\delta v(t) \sim \tau_L (\partial_t v_L) (t/\tau_L)
\label{eq:4.5.2}
\end{equation}
and therefore give the scaling exponents $\xi_p=p$.
This set of scaling exponent is trivially 
universal for any velocity field with $h>1/2$ and therefore 
a standard analysis of Lagrangian velocity fluctuations is unable
to disentangle the non trivial scaling component of the signal 
\cite{BBM_POF07}.

The statistical analysis of more than smooth signals has been recently 
addressed on the basis of an {\em exit-time statistics} \cite{BCLVV01}
in which one considers the time increments $T(\delta v)$ 
needed for a tracer to observe a change of $\delta v$ is its
velocity. 
Now, among the two contributions, in the limit of small
$\delta v(t)$, the differentiable part (\ref{eq:4.5.2}) 
will dominate except when the derivative $\partial_t v_L$ vanishes and the
local part (\ref{eq:4.5.1b}) becomes the leading one. 
For a signal with $1 \leq q \leq 2$,
its first derivative is a one-dimensional self-affine signal with
H\"older exponent $\xi = q-1$, which thus vanishes on a fractal set
of dimension $D=1-\xi=2-q$.

Therefore, the probability to observe the component $O(t^q)$ is
equal to the probability to pick a point on the fractal set of dimension $D$,
i.e.:
\begin{equation}
P(T \sim \delta v^{1/q}) \sim T^{1-D} \sim (\delta v)^{1-1/q} \; .
\label{eq:4.5.3}
\end{equation}
By using this probability for computing the average $p$-order moments
of exit-time statistics one obtains the following bi-fractal prediction
\cite{BCLVV01}
\begin{equation}
\langle T^{p}(\delta v) \rangle \sim \delta v^{\mu_p} \;, \;\;
\mbox{with} \;\; \mu_p=\min(p,\frac{p}{q}+1-\frac{1}{q}) \; .
\label{eq:4.5.4}
\end{equation}
According to prediction (\ref{eq:4.5.4}), low-order moments ($p \leq 1$) of the
inverse statistics only see the differentiable part of the signal,
while high-order moments ($p \geq 1$) are dominated by the local
fluctuations $O(t^q)$.

\begin{figure}[ht]
\includegraphics[width=10cm]{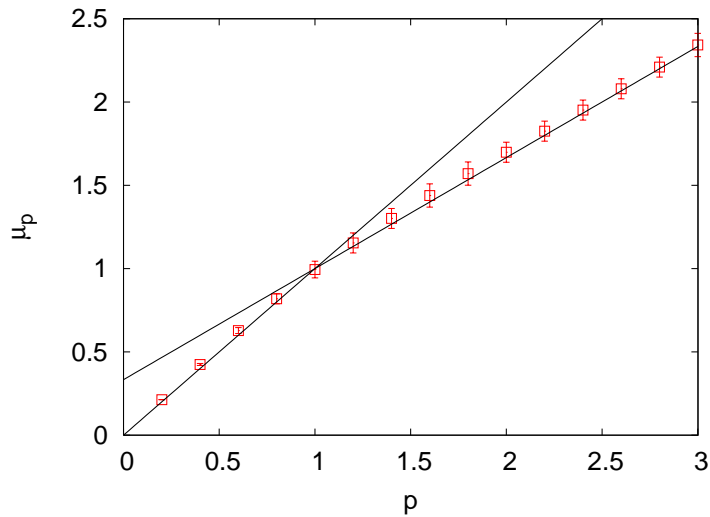}
\caption{The exit time scaling exponents $\mu_p$ 
from Lagrangian velocity fluctuations in a DNS of thermal
convection with Bolgiano-Obukhov scaling. 
Lines represent the bifractal prediction (\ref{eq:4.5.4})
and the error bars on the exponents have been estimated
by evaluating differences in $\mu_p$ changing the fitting interval.
Figure from \cite{BBM_POF07}.}
\label{fig:4.5.1}
\end{figure}

Figure~\ref{fig:4.5.1} shows the first moments of exit times
$\langle T^p(\delta v) \rangle$ computed from a direct numerical
simulation of two-dimensional Boussinesq equation forced by a mean 
(unstable) temperature gradient which generates an inverse cascade
with Bolgiano-Obukhov scaling. Particles are advected with
(\ref{eq:4.1.2}) and velocity fluctuations are collected along
Lagrangian trajectories. 
The bifractal spectrum predicted
by (\ref{eq:4.5.4}) is clearly reproduced. We remark that the
fact that for $p>1$ exponents follows the linear behavior
$\mu_p=(2 p +1)/3$ indicates the absence of intermittency
in Lagrangian statistics. This feature is a consequence of the
self-similarity of the inverse cascade in two-dimensional Bolgiano
convection.

%%%%%%%%%%%%%%%%%%%%%%%%%%%%%%%%%%%%%%%%%%%%%%%%%%%%%%%%%%%%
%%%%%%%%%%%%%%%%%%%%%%%%%%%%%%%%%%%%%%%%%%%%%%%%%%%%%%%%%%%%

%%%%%%%%%%%%%%%%%%%%%%%%%%%%%%%%%%%%%%%%%%%%%%%%%%%%%%%%%%%%
\section{Conclusions}
Starting from the seminal work of Kolmogorov we considered
the statistical features (mainly scaling properties),
both Eulerian and Lagrangian,
of the fully developed turbulence in the framework
of the multifractal model, i.e. in term of $D(h)$.
The hard, still unsolved, problem is, of course, to compute
$D(h)$ from first principles.
Up to now the unique doable approach is to use multiplicative models
motivated by phenomenological arguments.
The non trivial result is the fact that, once $D(h)$ is obtained with
a fit of the experimental data from the scaling exponents $\zeta_p$,
then  the other predictions obtained in the multifractal framework,
  e.g. the pdf of the velocity gradient,
the existence of an intermediate dissipative range, the scaling
of Lagrangian structure functions,  are well verified.

\subsection{Acknowledgments}
We are deeply grateful to Stefano Berti, Lamberto Rondoni and 
Massimo Vergassola for many useful remarks.
We thank Uriel Frisch and Mogens Jensen
for reviewing the manuscript with useful comments.
This work has been partially supported by PRIN 2005 project n.~2005027808 
and by CINFAI consortium (AM).

%%%%%%%%%%%%%%%%%%%%%%%%%%%%%%%%%%%%%%%%%%%%%%%%%%%%%%%%%%%%
\appendix
\section*{Appendix}
\setcounter{section}{1}
\appendix
\section{Synthetic turbulence: how to generate multiaffine 
stochastic processes}
In this Appendix we describe two methods for the
generation of multi-affine stochastic signals~\cite{BBCPVV93,BBCCV98},
whose scaling properties are fully under control. 
One is based on a dyadic decomposition
of the signal in a wavelet basis with a suitable assigned series of
stochastic coefficients \cite{BBCPVV93}. The second is based on a
multiplication of sequential Langevin-processes with a hierarchy of
different characteristic times \cite{BBCCV98}.  \\ The first procedure
is particularly appealing for modeling of spatial turbulent
fluctuations, because of the natural identification between wavelets
and eddies in the physical space. The second one looks more
appropriate for mimicking the turbulent time evolution in a fixed
point of the space.

Using the two methods it is possible
to build a rather general $(d+1)$-dimensional
process, $v({\bf x},t)$, with given scaling properties in time and in
space.

%%%%%%%%%%%%%%%%%%%%%%%%%%%%%%%%%%%%%%%%%%%%%%%%%%%%%%%%%%%%%%%%% 
\subsection{An algorithm based and dyadic decomposition}
\label{sec:synth.1}
%%%%%%%%%%%%%%%%%%%%%%%%%%%%%%%%%%%%%%%%%%%%%%%%%%%%%%%%%%%%%%%%% 
A non-sequential algorithm for $1$-dimensional multi-affine signal in
$[0,1]$, $v(x)$, can be introduced as~\cite{BBCPVV93}:
\begin{equation}
v(x) = \sum_{n=1}^N\sum_{k=1}^{2^{(n-1)}} a_{n,k}\,
                   \varphi\!\left(\frac{x-x_{n,k}}{\ell_n}\right)
\label{A.1}
\end{equation}
where we have a set of reference scales $\ell_n=2^{-n}$ and
$\varphi(x)$ is a wavelet-like function \cite{F92}, i.e. of zero mean
and rapidly decaying in both real space and Fourier-space.  The signal
$v(x)$ is built in terms of a superposition of fluctuations,
$\varphi((x-x_{n,k})/\ell_n)$ of characteristic width $\ell_n$ and
centered in different points of $[0,1]$, $x_{n,k} = (2k+1)/2^{n+1}$.
It has been proved \cite{BBCCV98}  that provided the coefficients
$a_{n,k}$ are chosen by a random multiplicative process, i.e.  the
daughter is given in terms of the mother by a random process,
$a_{n+1,k'} = X a_{n,k}$ with $X$ a random number identical,
independent distributed for any $\{n,k\}$, then the result of the
superposition is a multi-affine function with given scaling exponents,
namely:
$$
\langle |v(x+\ell)-v(x)|^p \rangle \sim
\ell^{\,\zeta_p}\,\,,
$$ 
with $\zeta_p= -p/2 - \log_2  \langle \langle X^p  \rangle \rangle$ 
and $ \ell_N \leq \ell \leq 1$.  
In this Appendix $\langle \langle \cdot \rangle \rangle$ 
indicates the
average over the probability distribution of the multiplicative
process.

Besides the rigorous proof, the rationale for the previous result is
simply that due to the hierarchical organization of the fluctuations,
one may easily see that the term dominating the expression of a
velocity fluctuation at scale $R$, in (\ref{A.1}) is given by the
couple of indices $\{n,k\}$ such that $n \sim log_2(R)$ and $x \sim
x_{n,k}$, i.e.  $v(x+\ell)-v(x) \sim a_{n,k}$.  The generalization
(\ref{A.1}) to d-dimension is given by:
$$
v({\bf x}) = \sum_{n=1}^N\sum_{k=1}^{2^{d(n-1)}} a_{n,k}\,
 \varphi\!\left(\frac{{\bf x}-{\bf x}_{n,k}}{\ell_n}\right)\,\,,
$$
where now the coefficients $\{a_{n,k}\}$ are given in terms of a
d-dimensional dyadic multiplicative process.  

\subsection{A sequential algorithm}
Sequential algorithms look more suitable for
mimicking temporal fluctuations. With the application to
time-fluctuations in mind, we will denote now the stochastic
1-dimensional functions with $u(t)$. The signal $u(t)$ is obtained by
a superposition of functions with different characteristic times,
representing eddies of various sizes~\cite{BBCCV98}:
\begin{equation}
u(t)=\sum_{n=1}^N u_n(t) \; .
\label{A.2}
\end{equation}
The functions $u_n(t)$ are defined by the multiplicative process
\begin{equation}
u_n(t)=g_n(t)x_1(t)x_2(t)\ldots x_n(t) \; ,
\label{A.3}
\end{equation}
where the $g_n(t)$ are independent stationary random processes, whose
correlation times are supposed to be $\tau_n=(\ell_n)^\alpha$, where
$\alpha = 1-h$ (i.e. $\tau_n$ are the eddy-turn-over time at scale
$\ell_n$) in the quasi-Lagrangian frame of reference \cite{LPP97} and
$\alpha = 1$ if one considers $u(t)$ as the time signal in a given
point, and $\langle g_n^2 \rangle = (\ell_n)^{2h}$, where $h$ is the
H\"older exponent. For a signal mimicking a turbulent flow, ignoring
intermittency, we would have $h=1/3$.  Scaling will appear for all
time delays larger than the UV cutoff $\tau_N$ and smaller than the IR
cutoff $\tau_1$.  The $x_j(t)$ are independent, positive defined,
identical distributed random processes whose time correlation decays
with the characteristic time $\tau_j$. The probability distribution of
$x_j$ determines the intermittency of the process.

The origin of (\ref{A.3}) is fairly clear in the context of fully
developed turbulence. Indeed we can identify $u_n$ with the velocity
difference at scale $\ell_n$ and $x_j$ with
$(\varepsilon_j/\varepsilon_{j-1})^{1/3}$, where $\varepsilon_j$ is
the energy dissipation at scale $\ell_j$ \cite{BBCCV98}.

The following arguments show, that the process defined according to
(\ref{A.2},\ref{A.3}) is multi-affine.  Because of the fast
decrease of the correlation times $\tau_j=(\ell_j)^\alpha$, the
characteristic time of $u_n(t)$ is of the order of the shortest one,
i.e., $\tau_n=(\ell_n)^\alpha$.  Therefore, the leading contribution
to the structure function $\tilde{S}_q(\tau) =
\langle |u(t+\tau)-u(t)|^q \rangle$ with $\tau \sim
\tau_n$ stems from the $n$-th term in (\ref{A.2}).  This can be
understood noting that in $u(t+\tau)-u(t) = \sum_{k=1}^N
[u_k(t+\tau)-u_k(t)]$ the terms with $k \le n$ are negligible because
$u_k(t+\tau) \simeq u_k(t)$ and the terms with $k \ge n$ are
sub-leading.  Thus one has:
\begin{equation}
\tilde{S}_q(\tau_n) 
\sim \langle |u_n|^q \rangle \sim 
 \langle \langle |g_n|^q \rangle \rangle 
 \langle \langle x^q  \rangle \rangle^n
\sim \tau_n^{\frac{h q}{\alpha} - \frac
{\log_2\langle\langle x^{q} \rangle \rangle}{\alpha}} 
\end{equation}
and therefore for the scaling exponents:
\begin{equation}
\zeta_q={h q \over \alpha} - 
{\log_2 \langle \langle x^{q} \rangle \rangle \over \alpha} \; .
\end{equation}
The limit of an affine function can be obtained when all the $x_j$ are
equal to $1$. A proper proof of these result can be found in
\cite{BBCCV98}.  \\ Let us notice at this stage that the previous
``temporal'' signal for $\alpha = 1 - h$ is a good candidate for a
velocity measurements in a Lagrangian, co-moving frame of reference
\cite{LPP97}.  Indeed, in such a reference frame the temporal
decorrelation properties at scale $\ell_n$ are given by the
eddy-turn-over times $\tau_n=(\ell_n)^{1-h}$.  On the other hand, in
the laboratory reference frame the sweeping dominates the time
evolution in a fixed point of the space and we must use as
characteristic times of the processes $x_n(t)$ the sweeping times
$\tau_n^{(s)} = \ell_n$, i.e., $\alpha=1$.


\section*{References}

\begin{thebibliography}{300}

\bibitem{PF85} 
Parisi G and Frisch U 1985 
in {\it Turbulence and predictability of geophysical fluid dynamics},
Eds Ghil M  Benzi R and Parisi G page 84 (Amsterdam: North-Holland)

\bibitem{BPPV84} 
Benzi R, Paladin G, Parisi G and Vulpiani A 1984
%``On the multifractal nature of fully developed turbulence and
%chaotic systems'',
{\it J. Phys. A: Math. Gen.} {\bf 17}  3521

\bibitem{Ellis99}
Ellis R S 1999
{\it Physica D} {\bf 133} 106

\bibitem{Varadhan03}
Varadhan S R S, in 
{\it Entropy} Eds Greve A  Keller  G and Warnecke D
page 199 (Princeton: Princeton Un. Press)


\bibitem{K62}
Kolmogorov A N 1962
{\it J. Fluid Mech.} {\bf 13} 82

\bibitem{NS64}
Novikov  E A and Stewart  R W 1964
{\it Izv. Akad. Nauk SSSR Geofiz.} {\bf 3} 408

\bibitem{M74}
Mandelbrot B B 1974
{\it J. Fluid Mech.} {\bf 62} 331

\bibitem{BS95}
Beck C and Sch\"ogl F 1995
{\it  Thermodynamics of Chaotic Systems}
(Cambridge: Cambridge University Press)

\bibitem{BP97}
Badii R and Politi A 1997
{\it Complexity: Hierarchical Structures and Scaling in Physics}
(Cambridge: Cambridge University Press)

\bibitem{Meakin98}
Meakin P 1998
{\it  Fractals, scaling and growth far from equilibrium}
(Cambridge: Cambridge University Press)

\bibitem{Harte01}
Harte D 2001
{\it Multifractals} (Boca Raton: Chapman and Hall/CRC)

\bibitem{HJKPS86}
Halsey T C, Jensen M H, Kadanoff L P, Procaccia I and Shraiman B I, 1986
%``''
{\it Phys. Rev. A} {\bf 33} 1141

\bibitem{F95} 
Frisch U 1995 
{\it Turbulence: the legacy of A. N. Kolmogorov}, 
(Cambridge: Cambridge University Press)

\bibitem{BJPV98} 
Bohr T, Jensen M H, Paladin G and Vulpiani A 1998 
{\it Dynamical systems approach to turbulence}, 
(Cambridge: Cambridge University Press)

\bibitem{MY75} 
Monin A and Yaglom A 1971 and 1975 
{\it Statistical Fluid Dynamics}, Vol. I and II
(Cambridge MA: MIT Press)

\bibitem{R22} 
Richardson L F 1922 
{\it Weather prediction by numerical processes}
(Cambridge: Cambridge University Press)

\bibitem {K41} 
Kolmogorov A N 1941 
%``The local structure of turbulence in incompressible viscous fluid 
%for very large Reynold number'', 
{\it Dokl. Akad. Nauk. SSSR} {\bf 30} 299;
reprinted in Kolmogorov A N 1991 
{\it Proc. R. Soc. Lond. A}  {\bf 434} 9. 

\bibitem{AGHA84} 
Anselmet F, Gagne Y, Hopfinger E J and Antonia R A 1984
%``High order velocity structure functions in turbulent shear flow'',
{\it J. Fluid. Mech.} {\bf 140}  63

\bibitem{PV87}
Paladin G and Vulpiani A 1987
{\it Phys. Rep.} {\bf 156} 147

\bibitem{SheLeveque94}
She Z S and L\'ev\^eque E 1994
{\it Phys. Rev. Lett.} {\bf 72} 336

\bibitem{AFLV92}
Aurell E  Frisch U  Lutsho J and Vergassola M 1992
{\it J. Fluid Mech.}  {\bf 238} 467 

\bibitem{BO99}
Bender C M and Orszag S A 1999 
{\it Advanced mathematical methods for scientists and engineers}, 
(New York: Springer)

\bibitem{FMMY05} 
Frisch U, Martins Afonso M, Mazzino A and Yakhot V 2005
%``Does multifractal theory of turbulence have logarithms in the scaling relations?'' 
{\it J. Fluid Mech.} {\bf 542} 97

\bibitem{MS89}
Meneveau C and  Sreenivasan K R 1989
%``Measurement of $f(\alpha )$ from scaling of histograms, and applications to dynamical systems and fully developed turbulence'',
{\it Phys. Lett. A} {\bf 137} 103 


\bibitem {vdWS88}
van de Water W and  Schram P 1988 
%``Generalized dimensions from near-neighbor information'', 
{\it Phys. Rev. A} {\bf 37} 3118

\bibitem{PV_PRA87}
Paladin G and Vulpiani A 1987
{\it Phys. Rev. A.} {\bf 35} 1971

\bibitem{FV91} 
Frisch U and Vergassola M 1991
%``A prediction of the multifractal model -- The Intermediate 
%Dissipation Range --'' 
{\it Europhys. Lett.} {\bf 14} 439

\bibitem{FS91}
Frisch U and  She Z -S 1991 
{\it Fluid Dynamic Research} {\bf 8} 139

\bibitem{CGH90} 
Castaing B, Gagne Y and  Hopfinger E J 1990
{\it Physica D} {\bf 46} 177 

\bibitem{VM91} 
Vincent A and  Meneguzzi M 1991
{\it J. Fluid Mech.} {\bf 225} 1 

\bibitem{BBPVV91} 
Benzi R, Biferale L,  Paladin G,  Vulpiani A and  Vergassola M 1991 
{\it Phys. Rev. Lett}  {\bf 67}  2299 

\bibitem{GC91}
Gagne Y and  Castaing B 1991
{\it C. R. Acad. Sci} Serie II {\bf 312} 441

\bibitem{BCVV99} 
Biferale L, Cencini M, Vergni D and  Vulpiani A 1999
%``Exit time of turbulent signals: a way to detect the Intermediate
%Dissipative Range'', 
{\it Phys. Rev. E} {\bf 60} R6295 


\bibitem {J99} 
Jensen M H 1999
%``Multiscaling and Structure Functions in Turbulence: 
%An Alternative Approach'', 
{\it Phys. Rev. Lett.} {\bf 83} 76

\bibitem{BCFV02}
Boffetta G, Cencini M, Falcioni M and Vulpiani A 2002
%``Predictability: a way to characterize Complexity''
Phys. Reports {\bf 356}, 367.

\bibitem{JPV91}
Jensen M H, Paladin G and  Vulpiani A 1991  
%``Multiscaling in Multifractals'' 
Phys. Rev. Lett.  {\bf 67} 208

\bibitem{ShraimanSiggia00}
Shraiman B I and Siggia E D 2000
{\it Nature} {\bf 405} 639

\bibitem{Lamb45}
Lamb H 1945
{\it Hydrodynamics}
(New York: New York Dover Publ.)

\bibitem{T88}
Tritton D J 1988
{\it Physical fluid dynamics},
(Oxford: Oxford Science Publ.)

\bibitem{MR83}
Maxey M R and Riley J J  1983 
{\it Phys. Fluids} {\bf 26} 883

%\bibitem{CFPrV90}
\bibitem{BBCLT06a}
Bec J, Biferale L, Cencini M, Lanotte A S and Toschi F 2006
{\it Phys. Fluids} {\bf 18} 081702

\bibitem{H66}
H\'enon M 1966
{\it C. R. Acad. Sci. Paris A} {\bf 262} 312

\bibitem{A84}
Aref H 1984
{\it J. Fluid Mech.} {\bf 143} 1

\bibitem{Chandrasekhar43}
Chandrasekhar S 1943
{\it Rev. Mod. Phys.} {\bf 15} 1

\bibitem{TL72}
Tennekes H and Lumley J L  1972
{\it A First Course in Turbulence}
(Cambridge MA: MIT Press)

\bibitem{Borgas93}
Borgas M S 1993 
{\it Phil. Trans. R. Soc. Lond. A} {\bf 342} 379

\bibitem{BDM02}
Boffetta G, De Lillo F and Musacchio S 2002
{\it Phys. Rev. E} {\bf 66} 066307

\bibitem{MMMP01}
Mordant N, Metz P, Michel O and Pinton J F  2001
{\it Phys. Rev. Lett.} {\bf 87} 214501

\bibitem{BBCLT06}
Biferale L, Boffetta G, Celani A, Lanotte A and Toschi F 2006
{\it J. Turbulence} {\bf 7} N6

\bibitem{BBCLT05}
Biferale L, Boffetta G, Celani A, Lanotte A and Toschi F 2005
{\it Phys. Fluids} {\bf 17} 021701

\bibitem{BCTBMS93}
Benzi R, Ciliberto S, Tripiccione R, Baudet C, Massaioli F and
Succi S 1993
{\it Phys. Rev. E} {\bf 48} R29

\bibitem{LVCAB01}
La Porta A, Voth G A, Crawford A M, Alexander J and Bodenschatz E
2001
{\it Nature} {\bf 409} 1017

\bibitem{BBCDLT04}
Biferale L, Boffetta G, Celani A, Devenish B J
Lanotte A and Toschi F 2004
{\it Phys. Rev. Lett.} {\bf 93} 064502


\bibitem{Richardson26}
Richardson L F 1926
{\it Proc. R. Soc. London A} {\bf 110} 709

\bibitem{Batchelor52}
Batchelor G K 1952
{\it Proc. Camb. Phil. Soc.} {\bf 48} 345

\bibitem{Obukhov41}
Obukhov A 1941
{\it Izv. Akad. SSSR. Ser. Geogr. Geofiz.} {\bf 5} 453

\bibitem{BS_POF02}
Boffetta G and Sokolov I M  2002
{\it Phys. Fluids} {\bf 14} 3224

\bibitem{BCCV_PRE99}
Boffetta G, Celani A, Crisanti A and Vulpiani A 1999
{\it Phys. Rev. E} {\bf 60} 6734

\bibitem{ABCCV97}
Artale V, Boffetta G, Celani A, Cencini M and Vulpiani A 1997
{\it Phys. Fluids A} {\bf 9} 3162

\bibitem{BS_PRL02}
Boffetta G and Sokolov I M 2002
{\it Phys. Rev. Lett.} {\bf 88} 094501

\bibitem{BBCDLT_POF05}
Biferale L, Boffetta G, Celani A, Devenish B J, Lanotte A
and Toschi F 2005
{\it Phys. Fluids} {\bf 17} 115101

\bibitem{BBM_POF07}
Bistagnino A, Boffetta G and Mazzino A 2007
{\it Phys. Fluids} {\bf 19} 011703

\bibitem{OM_JFM00}
Ott S and Mann J 2000
{\it J. Fluid Mech.} {\bf 422} 207


\bibitem{Siggia94}
Siggia E D 1994
{\it Annu. Rev. Fluid Mech.} {\bf 26} 137

\bibitem{Chertkov03}
Chertkov M 2003
{\it Phys. Rev. Lett.} {\bf 91}, 115001

\bibitem{ZWX05}
Zhang J, Wu X L and Xia K Q  2005 
{\it Phys. Rev. Lett.} {\bf 94} 174503

\bibitem{CMV01}
Celani A, Mazzino A and Vergassola M 2001
{\it Phys. Fluids} {\bf 13} 2133


\bibitem{BCLVV01}
Biferale L, Cencini M, Lanotte A, Vergni D and Vulpiani A 2001
{\it Phys. Rev. Lett.} {\bf 87} 124501

\bibitem{BBCPVV93} 
Benzi R, Biferale L, Crisanti A, Paladin G, Vergassola M and 
Vulpiani A 1993
%``A Random process for the construction of multiaffine fields'',
{\it Physica D} {\bf 65} 352


\bibitem{BBCCV98} 
Biferale L, Boffetta G, Celani A, Crisanti A and Vulpiani A 1998
%``Mimicking a turbulent signal: sequential multiaffine processes'',
{\it Phys. Rev. E} {\bf 57} R6261


\bibitem{F92} 
Farge M 1992 
%``Wavelet transforms and their applications to turbulence'',
{\it Ann. Rev. Fluid Mech.} {\bf 24} 395
 
\bibitem{LPP97} 
L'vov V S, Podivilov E and  Procaccia I 1997
%``Temporal multiscaling in hydrodynamic turbulence'',
{\em Phys. Rev. E}  {\bf 55} 7030







%\bibitem {J99} 
%Jensen M H 1999
%%``Multiscaling and Structure Functions in Turbulence: 
%%An Alternative Approach'', 
%{\it Phys. Rev. Lett.} {\bf 83} 76
%

%\bibitem{M75} 
%Mandelbrot B  1975
%%``On the geometry of homogeneous turbulence, with stress on the fractal 
%%dimension of the iso-surfaces of scalars'',
%{\it J. Fluid. Mech.} {\bf 72} 401
%
%\bibitem {PSV95}
%Paladin G, Serva M and Vulpiani A 1995 
%%``Complexity in dynamical systems with noise'', 
%{\it Phys. Rev. Lett.} {\bf 74} 66
























\end{thebibliography}
\end{document}